\documentclass[11pt,epsf]{article}
\usepackage{amsmath,amssymb}
\usepackage{graphicx}
\usepackage{cite,fancyhdr}
\usepackage{url}

\usepackage[affil-it]{authblk}
\usepackage[left=2.5cm,right=2.5cm,top=2.5cm,bottom=2.5cm,a4paper]{geometry}

\allowdisplaybreaks

\begin{document}

\setcounter{footnote}{0}
\setcounter{figure}{0}
\setcounter{table}{0}

\title{\bf \Large 
Lepton flavor violations in SUSY models for muon $g-2$ with right-handed neutrinos
}
\author[1,2]{\normalsize Ryo Nagai}
\author[3]{\normalsize Norimi Yokozaki}

\affil[1]{\small 
Dipartimento di Fisica e Astronomia, Universita' degli Studi di Padova, Via Marzolo 8, 35131 Padova, Italy
}
\affil[2]{\small 
Istituto Nazionale di Fisica Nucleare (INFN), Sezione di Padova, Via Marzolo 8, 35131 Padova, Italy
}
\affil[3]{\small 
KEK Theory Center, IPNS, KEK, Tsukuba, Ibaraki 305–0801, Japan}

\date{}

\maketitle

\thispagestyle{fancy}
\rhead{KEK-TH-2235}
\cfoot{\thepage}
\renewcommand{\headrulewidth}{0pt}

\begin{abstract}
\noindent
We consider supersymmetric (SUSY) models for the muon $g-2$ anomaly 
without flavor violating masses at the tree-level. The models can avoid LHC constraints and the vacuum stability constraint in the stau-Higgs potential. Although large flavor violating processes are not induced within the framework of minimal SUSY standard model, once we adopt a seesaw model, sizable lepton flavor violating (LFV) processes such as $\mu \to e \gamma$ and $\mu \to e$ conversion are induced. These LFV processes will be observed at future experiments such as MEG-II, COMET and Mu2e if right-handed neutrinos are heavier than $10^9$\,GeV motivated by the successful leptogenesis. This conclusion is somewhat model independent since Higgs doublets are required to have large soft SUSY breaking masses, leading to flavor violations in a slepton sector via neutrino Yukawa interactions.
\end{abstract}

\clearpage

\section{Introduction}
The supersymmetric (SUSY) standard model is one of the most attractive candidates for new physics beyond the Standard model (SM). In the minimal SUSY standard model (MSSM), SM gauge couplings beautifully unify at the scale around $10^{16}$ GeV. This indicates the existence of a grand unified theory (GUT), which naturally explains the charge quantization. We now have dark matter (DM) candidates in MSSM. The large hierarchy between the electroweak symmetry breaking (EWSB) scale and Planck/GUT scale is stabilized due to the absence of quadratic divergences. Moreover, interestingly, the long standing anomaly of the muon $g-2$~\cite{Bennett:2006fi, Campanario:2019mjh,  Davier:2019can, Keshavarzi:2019abf,g-2summary} is explained if masses of smuons and electroweak gauginos are $\mathcal{O}(100)$\,GeV~\cite{Moroi:1995yh}. The situation of the muon $g-2$ anomaly is expected to become clearer near future~\cite{Keshavarzi:2019bjn} (see also~\cite{Mibe:2011zz, Abe:2019thb}).

The lightness of sleptons and electroweak gauginos generally leads to unacceptably large flavor violating processes such as $\mu \to e\gamma$ and $\mu \to e$ conversion. These flavor violating processes originate from soft SUSY breaking mass parameters which mixes different generations of sfermions.
The dangerous flavor violating sfermion masses are avoided when the SUSY breaking masses are generated through gauge interactions and SM Yukawa interactions, leading us to gaugino mediation~\cite{Inoue:1991rk, Kaplan:1999ac, Chacko:1999mi} or Higgs mediation~\cite{Yamaguchi:2016oqz,Yin:2016shg}.\footnote{
Gauge mediation models are also viable options for the muon $g-2$ while avoiding the too large flavor violating processes. See e.g. Refs.~\cite{Ibe:2012qu, Bhattacharyya:2018inr}.
} In these mediation mechanisms, the slepton and squark masses vanish at the tree-level and they are generated radiatively via gaugino loops or Higgs loops. 
Therefore, the flavor problem is absent within MSSM even if some SUSY particles are as light as $\mathcal{O}(0.1 \mathchar`-1\,{\rm TeV})$~\cite{Yanagida:2018eho}.

Another important constraint on the models with the light sleptons is vacuum stability constraint in the stau-Higgs potential: if $\mu \tan\beta$ is too large, 
the EWSB minimum decays to a charge breaking minimum, where the staus have non-zero vacuum expectation values (VEVs), 
with a too short life time~\cite{Rattazzi:1996fb,Hisano:2010re,Kitahara:2013lfa,Endo:2013lva,Chowdhury:2013dka}. Here, $\mu$ is a Higgsino mass parameter and $\tan\beta$ is a ratio of the VEVs of Higgs doublets. 
This constraint is avoided if the staus are (much) heavier than smuons or $\mu$ is not large, 
which requires large soft SUSY breaking masses for the Higgs doublets as will be shown later.

In Refs.~\cite{Harigaya:2015kfa, Cox:2018vsv}, it has been shown that, in realistic ultraviolet (UV) models of gaugino and Higgs mediation, the muon $g-2$ anomaly is completely solved within MSSM while avoiding stringent LHC constraints and the vacuum stability constraint.
In particular, the model of Higgs mediation with non-universal gaugino masses can also explain the correct relic abundance of dark matter without conflicting direct and indirect experiments~\cite{Cox:2018vsv}.\footnote{This model can be regarded as a modification of Higgs-anomaly mediation presented in Refs.~\cite{Yin:2016shg,Yanagida:2016kag,Yanagida:2020jzy}} 
See also, e.g., Refs.~\cite{Endo:2017zrj,Cox:2018qyi,Datta:2018lup,Tran:2018kxv,Badziak:2019gaf,Abdughani:2019wai,Endo:2020mqz,Chakraborti:2020vjp} for recent studies based on phenomenological models explaining the muon $g-2$.

In this paper, we extend the previous studies by including three right-handed (RH) neutrinos,\footnote{
There are studies of lepton flavor violations in high scale SUSY models with RH neutrinos of $\sim 10^{15}$\,GeV~\cite{Moroi:2013vya,Evans:2018ewb}. In these studies, all the sfermions including the smuons are heavier than $\sim 10$\,TeV.} which enables us to explain the tiny neutrino masses via the seesaw mechanism~\cite{Yanagida:1979as,GellMann:1980vs} (see also Ref.~\cite{Minkowski:1977sc}). 
With the inclusion of the RH neutrinos, flavor violating elements of the left-handed (LH) slepton mass matrix are induced by a renormalization group (RG) running effect~\cite{Hisano:1995nq} even in the models of gaugino and Higgs mediation. This is because 
the soft SUSY breaking mass for the up-type Higgs is non-vanishing and large at the tree-level, and the up-type Higgs couples to chiral multiplets of LH leptons and RH neutrinos through flavor violating neutrino Yukawa couplings. Consequently, lepton flavor violating (LFV) processes such as $\mu \to e \gamma$ and $\mu \to e$ conversion become non-negligible and detectable at future LFV experiments 
when the RH neutrinos are heavier than $\mathcal{O}(10^9)$\,GeV.\footnote{If RH neutrinos are as light as $\sim1$\,TeV and the neutrino Yukawa couplings are large, 
loop diagrams involving RH neutrinos contribute to the muon $g-2$ and LFVs violations (see e.g. Refs.~\cite{Ilakovac:2012sh, Ilakovac:2013wfa}). 
In this case, the small neutrino masses can be explained by assuming a special form of the neutrino Yukawa matrix.} In fact, the RH neutrinos heavier than $\mathcal{O}(10^9)$\,GeV are motivated by the successful thermal leptogenesis~\cite{Fukugita:1986hr,Davidson:2002qv} for the baryon asymmetry of the universe.

\section{SUSY models for muon $g-2$}
We introduce the three different models for the muon $g-2$ 
without flavor violating masses at the tree-level.
All of the models include the direct couplings between the Higgs fields and a SUSY breaking field $Z$, which are needed to avoid the vacuum stability constraint in the stau-Higgs potential. 

\subsection{CP-safe gaugino mediation model (model A)}
We first consider a gaugino mediation model given in Ref.~\cite{Harigaya:2015kfa}, which respects the shift symmetry of the SUSY breaking field $Z$: $Z \to Z + i \mathcal{R}$ with $\mathcal{R}$ being a real constant. With the shift symmetry, dangerous CP violating phases are suppressed~\cite{Iwamoto:2014ywa}. We refer to this model as the model {\bf A}. The K\"{a}hler potential is given by
\begin{eqnarray}
K = -3 \ln \left(1-\frac{f(x) + \phi_I^\dag \phi_I + H_u^\dag H_u + H_d^\dag H_d + \Delta K}{3} \right),
\end{eqnarray}
where $x=Z+Z^\dag$, $f(x)$ is an arbitrary function of $x$ and 
\begin{eqnarray}
\Delta K = g_u(x) H_u^\dag H_u + g_d(x) H_d^\dag H_d.
\end{eqnarray}
Here, $\Phi_I$ is a matter multiplet; the matter multiplets include three generations of leptons, quarks and RH neutrinos; $g_{u}(x)$ and $g_d(x)$ are arbitrary functions of $x$; we have omitted gauge interactions and taken $M_P=1$, where $M_P$ is the reduced Planck mass.
We assume the above K\"{a}hler potential is defined at the GUT scale. With the K\"{a}hler potential, all the sfermions are massless at the tree-level, which is a very important assumption to solve the SUSY flavor problem. The sfermions masses are dominantly generated from gaugino masses through radiative corrections (gaugino mediation).   

The superpotential is
\begin{eqnarray}
W = \mathcal{C} +  \mu H_u H_d + W_{\rm Yukawas} + W_{\rm RN},
\end{eqnarray}
where 
\begin{eqnarray}
W_{\rm RN} = \bar N_i (Y_\nu)_{ij} L_j H_u - \frac{1}{2} \bar{N}_i M_{N_i} \bar{N}_i, \label{eq:ynu}
\end{eqnarray}
and
\begin{eqnarray}
W_{\rm Yukawas} =  -\bar{U}_i (Y_u)_{ij} Q_j H_u + \bar{D}_i (Y_d)_i Q_i H_d   + \bar{E}_i (Y_e)_i L_i H_d.
\end{eqnarray}
Here, we have taken $M_N$, $Y_d$ and $Y_e$ to be diagonal by the field redefinitions of $\bar{N}_i$, $Q_i$, $\bar{D}_i$, $L_i$ and $\bar{E}_i$ without loss of generality.\footnote{
If the sfermion masses are not universal nor vanishing, the sfermion mass matrices change with the field redefinitions, inducing flavor mixings. These flavor mixings generally induce too large flavor changing processes.  
} The Yukawa coupling, $Y_u$, is given by
\begin{eqnarray}
Y_u = {\rm diag}(m_u, m_c, m_t)/\left<H_u\right> \times V_{\rm CKM},
\end{eqnarray}
with $V_{\rm CKM}$ being the CKM matrix. The mass parameter $\mathcal{C}$ is a constant term. 
We take $\mu$ and $\mathcal{C}$ to be real by $U(1)_R$ rotation and field redefinitions of $H_u$ and $H_d$.

The cosmological constant vanishes under the following condition:
\begin{eqnarray}
\left<\frac{\partial K}{\partial x}\right>^2 = 3 \left<\frac{\partial^2 K}{\partial x^2}\right> \to \left<\frac{\partial^2 f}{\partial x^2}\right> =0.
\end{eqnarray}
The SUSY is broken at the minimum of vanishing cosmological constant~\cite{Izawa:2010ym} and $F$-term of $Z$ is given by
\begin{eqnarray}
F_Z = - e^{\left<K\right>/2} 3 (1-\left<f\right>/3) \left<\frac{\partial f}{\partial x}\right>^{-1} \mathcal{C} = 
-3 n \times m_{3/2},
\end{eqnarray}
where $m_{3/2} = e^{\left<K\right>/2} \mathcal{C}$ is a gravitino mass and 
\begin{eqnarray}
n = \left< \frac{\partial f}{\partial x} \right>^{-1} (1-\left<f\right>/3).
\end{eqnarray}
Note that $F_Z$ is a real number since $n$ and $m_{3/2}$ are both real.

The canonically normalized kinetic terms for $\phi_I$, $H_u$ and $H_d$ are obtained by the following field redefinitions:
\begin{eqnarray}
 \phi_I &\to& (1-\left<f\right>/3)^{1/2} \phi_I , \nonumber \\
H_u &\to& \left[\frac{1+\left<g_u\right>}{1-\left<f\right>/3}\right]^{-1/2} H_u, \nonumber \\
 H_d &\to& \left[\frac{1+\left<g_d\right>}{1-\left<f\right>/3}\right]^{-1/2}H_d.
\end{eqnarray}
Accordingly, the parameters in the superpotential are rescaled as~\cite{wess_bagger}
\begin{eqnarray}
 \mu &\to& e^{-\left<K\right>/2} \left[\frac{1+\left<g_u\right>}{1-\left<f\right>/3}\right]^{1/2} \left[\frac{1+\left<g_d\right>}{1-\left<f\right>/3}\right]^{1/2} \mu, \nonumber \\
 Y_{u,\nu} &\to& e^{-\left<K\right>/2} (1+\left<g_u\right>)^{1/2} (1-\left<f\right>/3)^{-3/2} Y_{u,\nu}, \nonumber \\
Y_{d, e} &\to& e^{-\left<K\right>/2} (1+\left<g_d\right>)^{1/2} (1-\left<f\right>/3)^{-3/2} Y_{d,e}.
\end{eqnarray}

The soft SUSY breaking masses for $H_u$ and $H_d$ are 
\begin{eqnarray}
m_{H_u}^2 &=& 9 n^2  (c_u^2 -d_u) m_{3/2}^2, \nonumber \\
m_{H_d}^2 &=& 9 n^2  (c_d^2 - d_d) m_{3/2}^2,
\end{eqnarray}
where 
\begin{eqnarray}
c_u &=& \left<\frac{\partial g_u}{\partial x}\right>(1 + \left<g_u\right>)^{-1} , \ 
c_d = \left< \frac{\partial g_d}{\partial x}\right> (1 + \left<g_d\right>)^{-1},  \nonumber \\
d_u &=& \left<\frac{\partial^2 g_u}{\partial x^2}\right> (1 +\left< g_u\right>)^{-1}, \ 
d_d = \left<\frac{\partial^2 g_u}{\partial x^2}\right> (1 + \left<g_d\right>)^{-1} ,
\end{eqnarray}
with $c_u$, $c_d$, $d_u$ and $d_d$ being real numbers.

A-terms and the Higgs B-term are 
\begin{eqnarray}
A_{u} = A_{\nu} &=& -3 n c_u \times m_{3/2}, \nonumber \\
A_{d}= A_{e} &=& -3 n c_d \times m_{3/2}, \nonumber \\
B_\mu &=& (A_{u} + A_{d}) \mu,
\end{eqnarray}
where we have no CP violating phase.

The gaugino masses are generated by the coupling between $Z$ and field strength superfields in a way consistent with a grand unified theory. Here, we consider $SU(5) \times SU(3)_H \times U(1)_H$ product group unification~\cite{Yanagida:1994vq,Hotta:1995cd}, which solves the doublet triplet splitting problem in a simple way. 
The relevant couplings are
\begin{eqnarray}
  \mathcal{L} &=&  \int d^2\theta \left( \frac{1}{4 g_5^2} -  \frac{k_5 Z}{2} \right) W_5 W_5 + h.c. \nonumber \\
  &+& \int d^2\theta \left( \frac{1}{4 g_{3H}^2}  - \frac{k_{3H} Z}{2} \right) W_{3H} W_{3H}+ h.c. \nonumber \\
  &+&  \int d^2\theta \left( \frac{1}{4 g_{1H}^2} - \frac{k_{1H} Z}{2} \right) W_{1H} W_{1H} + h.c, \label{eq:gmasses}
\end{eqnarray}
where $g_5$, $g_{3H}$ and $g_{1H}$ are gauge couplings of $SU(5)$, $SU(3)_H$ and $U(1)_H$, respectively; 
$W_5$, $W_{3H}$ and $W_{1H}$ are the field strength superfields of $SU(5)$, $SU(3)_H$ and $U(1)_H$. 
Note that $k_5$, $k_{3H}$ and $k_{1H}$ are real respecting the shift symmetry so that no CP violating phases arise from the gaugino masses. 

After $SU(5) \times SU(3)_H \times U(1)_H$ is broken down to the SM gauge group, the gaugino masses are obtained as
\begin{eqnarray}
M_{1}&=& (k_5 \mathcal{N} + k_{1H}) \frac{g_5^2 g_{1H}^2}{g_5^2 + \mathcal{N} g_{1H}^2 } (-3 n ) m_{3/2},\nonumber \\
M_{2} &=& k_5 g_5^2 (-3 n ) m_{3/2},\nonumber \\
M_{3} &=& (k_5 + k_{3H}) \frac{g_5^2 g_{3H}^2}{g_5^2 + g_{3H}^2} (-3 n) m_{3/2},
\end{eqnarray}
where 
$M_{1}$, $M_{2}$ and $M_{3}$ are the bino, wino and gluino masses, respectively;
$\mathcal{N}$ is a real constant depending on the $U(1)_H$ charge of the GUT breaking Higgs field; 
we have rescaled the gauge couplings as $g_a^{-2}  \to g_a^{-2} + 2 k_a \left<Z\right>$.
We note that the gaugino masses are non-universal at the GUT scale, which is important to explain the muon $g-2$ while avoiding LHC constraints on colored SUSY particles~\cite{ATLAS-CONF-2019-040,Sirunyan:2019ctn}. 

Apart from the RH neutrino masses, the free parameters in this model are $m_{H_u}^2$, $m_{H_d}^2$, $A_u$, $B_\mu$, $M_1$, $M_2$ and $M_3$, or more conveniently, we choose 
\begin{eqnarray}
\mu,\, m_A,\, \tan\beta,\, A_u,\, M_1,\, M_2,\, M_3,
\end{eqnarray}
where $m_A$ is a CP-odd Higgs mass. Here, $\mu$, $m_A$ and $\tan\beta$ are defined at the EWSB scale while $M_1$, $M_2$ and $M_3$ are given at the GUT scale, $M_{\rm GUT}$.

In order to avoid the vacuum stability constraint in the stau-Higgs potential, we consider a small $\mu$ case. Then, the chargino contribution to the muon $g-2$ is dominant. The small $\mu$ is achieved by taking $m_{H_u}^2$ at the GUT scale to be large and positive. As we will see, this $m_{H_u}^2$ induces flavor violating slepton masses through the Yukawa interactions in Eq.~\eqref{eq:ynu}. 


\subsection{Higgs mediation model with bino-wino coannihilation (model B)}
Next, we consider a Higgs mediation model presented in Ref.~\cite{Cox:2018vsv}, focusing on the bino-wino coannihilation region~\cite{Baer:2005jq}, where the bino and wino masses are quasi degenerated at the EWSB scale.  
We refer to this model as the model {\bf B}. In this model, the Higgs soft masses are assumed to be tachyonic and large as $\mathcal{O}(10)$\,TeV. These Higgs soft masses lead to natural spitting of sfermion masses through radiative corrections~\cite{Yamaguchi:2016oqz,Yin:2016shg}: third generation sfermions become much heavier than first/second generation sfermions without inducing too large flavor violating masses~\cite{Yanagida:2018eho}. Then, the vacuum stability constraint in the stau-Higgs potential is easily avoided due to the heavy staus as discussed in Sec.~\ref{sec:results}. 

The K\"{a}hler potential is given by
\begin{eqnarray}
K' = -3 \ln \left(1-\frac{f'(Z, Z^\dag) + \phi_I^\dag \phi_I + H_u^\dag H_u + H_d^\dag H_d + \Delta K'}{3} \right),
\end{eqnarray}
where $f'(Z,Z^\dag)$ is a function of $Z$ and $Z^\dag$ and 
\begin{eqnarray}
\Delta K' = c_h Z^\dag Z( H_u^\dag H_u + \kappa_d H_d^\dag H_d) - (c_b |Z|^2 H_u H_d + h.c.). \label{eq:kahlerB}
\end{eqnarray}
Here, $c_h$ is assumed to be positive and $\kappa_d=1$. The concrete models justifying these assumptions are given in Ref.~\cite{Yanagida:2020jzy}. In this model, we do not consider the shift symmetry of $Z$. 
However, to construct a model with the shift symmetry is not difficult.

The superpotential is given by
\begin{eqnarray}
W' = \mathcal{C} +  w(z) + \mu H_u H_d + W_{\rm Yukawas} + W_{\rm RN}. \label{eq:superB}
\end{eqnarray}
By assuming $\left<Z\right> \simeq 0$, the SUSY breaking $F$-term is obtained as
\begin{eqnarray}
 \left<F_Z\right> \simeq - \left<\frac{\partial w(z)}{\partial Z}\right>^*,
\end{eqnarray}
where $|\left<F_Z\right>|^2 = 3 m_{3/2}^2$. Here, we take a canonically normalized kinetic term for $Z$.

From Eqs.~\eqref{eq:kahlerB} and \eqref{eq:superB}, we obtain
\begin{eqnarray}
 m_{H_u}^2 &=& m_{H_d}^2 = - 3 c_h m_{3/2}^2, \nonumber \\
 A_u&=&A_d=A_e=A_\nu=0, \nonumber \\
 B_\mu &=& 3 c_b m_{3/2}^2 - m_{3/2} \mu.
\end{eqnarray}

The gaugino masses are generated from the couplings between $Z$ and field strength superfields in Eq.~\eqref{eq:gmasses}, and they are non-universal at the GUT scale. 
This allows us to explain the correct relic abundance of dark matter through the bino-wino coannihilation, avoiding experimental constraints. The free parameters in this model are 
\begin{eqnarray}
 m_{H_u}^2,\, \tan\beta,\, M_1,\, M_2,\, M_3,\, {\rm sign}(\mu),
\end{eqnarray}
which are given at $M_{\rm GUT}$. In the following analysis, we take $\mu>0$.

\subsection{Higgs mediation model with bino-slepton coannihilation (model C)}
Lastly, we consider the Higgs mediation model focusing on the bino-slepton coannihilation region~\cite{Ellis:1999mm}, where the masses of the bino, selectron and smuon are quasi degenerated at the EWSB scale. 
We refer to this model as the model {\bf C} although the Lagrangian is completely same as that of the model {\bf B}. The only differences are as follows: the wino mass is larger and the Higgs soft masses are smaller compared to the model {\bf B}. The free parameters in this model are same as those in the model {\bf B}. 
We also take $\mu>0$.

\section{Lepton flavor violations and muon $g-2$} \label{sec:results}
In this section, we calculate the LFV processes in the model {\bf A}, {\bf B} and {\bf C}, focusing on parameter regions consistent with the muon $g-2$ experiment. The experimental value of the muon $g-2$~\cite{Bennett:2006fi} is deviated from a SM prediction (see \cite{g-2summary} and references therein) with a significance of 3.7\,$\sigma$ level:
\begin{eqnarray}
\Delta a_\mu = (279 \pm 76)\times 10^{-11}.
\end{eqnarray}
This deviation is explained only when the smuon(s) and electroweak gauginos are light as $O(100)$\,GeV together with a large $\tan\beta$ of $\mathcal{O}(10)$. In this case, the vacuum stability constraint in the stau-Higgs potential becomes important: if $\mu\tan\beta$ is too large, the EWSB minimum decays into the charge breaking minimum with a life time shorter than the age of the universe. The constraint is shown in Ref.~\cite{Endo:2013lva} as
\begin{eqnarray}
 \eta_{\tau}^{-1} \left|\frac{m_{\tau} \mu \tan\beta}{1+\Delta_\tau}\right| &\leq& 1.01\times 10^2 {\rm GeV}
 \sqrt{m_{\tilde{L}_3} m_{\tilde{E}_3}}
 +1.01\times10^2 {\rm GeV}(m_{\tilde{L}_3} +1.03 m_{\tilde{E}_3})
  \nonumber \\
&-&  2.27\times10^4{\rm GeV}^2 + \frac{2.97\times10^6{\rm GeV}^3}{m_{\tilde{L}_3} + m_{\tilde{E}_3}}
 -1.14\times 10^8{\rm GeV}^4
 \left(
 \frac{1}{m_{\tilde{L}_3}^2}
 +\frac{0.983}{m_{\tilde{E}_3}^2}
 \right), \nonumber \\ \label{eq:vac}
\end{eqnarray}
where $\Delta_\tau$ is a radiative correction to the tau Yukawa coupling~\cite{Guasch:2001wv}, and $m_{\tilde{L}_3} (m_{\tilde{E}_3})$ is the mass of LH (RH) stau. The normalization factor $\eta_{\tau} \approx 1$ slightly depends on $\tan\beta$~\cite{Kitahara:2013lfa}.
Clearly, this constraint is avoided when $\mu$ is small or staus are much heavier than the smuons, which requires large soft SUSY breaking masses for the Higgs doublets. 

Let us firstly consider the small $\mu$ case, where the chargino diagram dominantly contributes to the muon $g-2$. 
The $\mu$ parameter is determined by the EWSB condition, which is given by
\begin{eqnarray}
 \frac{m_Z^2}{2} &\simeq& - (m_{H_u}^2(M_{\rm GUT}) + \Delta m_{H_u}^2 + \mu^2) \nonumber \\
 &+& (m_{H_d}^2(M_{\rm GUT})+\Delta m_{H_d}^2 - m_{H_u}^2(M_{\rm GUT})-\Delta m_{H_u}^2)/\tan^2\beta + \dots,
\end{eqnarray}
where $\dots$ denotes higher order terms of $1/\tan^n\beta$ $(n \ge 4)$; 
$m_{H_u}(M_{\rm GUT})$ and $m_{H_d}(M_{\rm GUT})$ are soft SUSY breaking masses for $H_u$ and $H_d$, respectively;  $\Delta m_{H_u}^2$ and $\Delta m_{H_d}^2$ are radiative corrections. 
To explain the Higgs boson mass of 125\,{\rm GeV}, we need a large stop mass $m_{\tilde{t}}$ or a large trilinear coupling $A_t$~\cite{Okada:1990vk, Ellis:1990nz, Haber:1990aw, Okada:1990gg, Ellis:1991zd}. In this case, $\Delta m_{H_u}^2 \sim (m_{\tilde t}^2\  {\rm or}\  A_t^2)$, is inevitably large. Therefore, the small $\mu$ is only achieved with $m_{H_u}^2(M_{\rm GUT}) \sim (m_{\tilde t}^2\  {\rm or}\  A_t^2)$. Numerically, we find $m_{H_u}^2(M_{\rm GUT}) \sim 4\,{\rm TeV}^2$ to be consistent with $\mu \sim 100$\,GeV.

For the large $\mu$ case, where the neutralino diagram dominantly contributes to the muon $g-2$, we need the large stau masses to avoid the constraint in Eq.~\eqref{eq:vac}. 
One possibility is to make the staus heavy by hand as explored in Refs.~\cite{Ibe:2013oha,Ibe:2019jbx,Han:2020exx}. 
However, in this case, the constraint from $\mu \to e \gamma$ is too severe unless we assume a special structure of the lepton Yukawa couplings~\cite{Ibe:2019jbx}. 
Alternatively, we can make the staus heavy using the Higgs-loop effects~\cite{Yamaguchi:2016oqz,Yin:2016shg}, without inducing LFV in the framework of MSSM. Here, the Higgs soft masses are assumed to be large as $m_{H_{u,d}}=\mathcal{O}(10)$\,TeV and tachyonic. Then, the staus become heavy as $\sim 10$\,TeV by a Higgs loop at the one-loop level due to the large tau-Yukawa coupling while the selectrons and smuons remain light as $\mathcal{O}(100)$\,GeV.\footnote{The selectron and smuon masses are dominantly generated by two-loop diagrams involving the Higgs doublets and one-loop diagrams involving gauginos.} The generated stau masses are estimated as
\begin{eqnarray}
 m_{\tilde{L}_3}^2 = (m_{\tilde{L}}^2)_{33} \sim \frac{Y_{\tau}^2}{8\pi^2} |m_{H_u}^2| \ln \frac{M_{\rm GUT}}{M_{\rm SUSY}},
\label{eq:mL3}
\end{eqnarray}
and
\begin{eqnarray}
 m_{\tilde{E}_3}^2 = (m_{\tilde{E}}^2)_{33} \sim \frac{Y_{\tau}^2}{4\pi^2} |m_{H_u}^2| \ln \frac{M_{\rm GUT}}{M_{\rm SUSY}},
\end{eqnarray}
where $m_{H_d}^2 \sim m_{H_u}^2$ is used and $M_{\rm SUSY}$ is a SUSY particle mass scale. We note that the condition, $m_{H_d}^2 \sim m_{H_u}^2$, is required so that a $U(1)_Y$ $D$-term contribution to the sfermion masses proportional to $(m_{H_u}^2-m_{H_d}^2)$ is not too large: if the $D$-term contribution is too large, the smuon and slectron become tachyonic. See appendix \ref{sec:app_a} for renormalization group equations for the slepton masses.

\begin{figure}[!t]
\begin{center}  
\includegraphics[width=78mm]{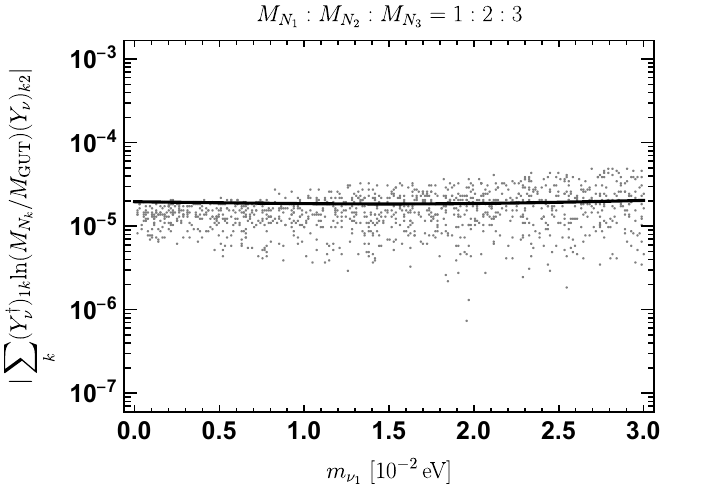}
\includegraphics[width=78mm]{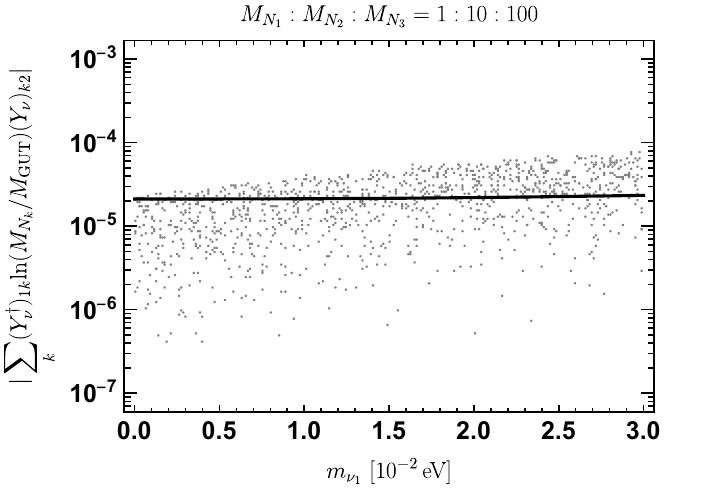}
\end{center}
\caption{
The $m_{\nu_1}$ dependence in the LFV coupling parameter. We fix $\tan\beta=20$ and $M_{N_3}=10^{10}\,\mbox{GeV}$ and we consider two cases: i) $M_{N_1}:M_{N_2}:M_{N_3}=1:2:3$ and ii) $M_{N_1}:M_{N_2}:M_{N_3}=1:10:100$. The black line corresponds to the case with taking $R=1$.
}
\label{fig:nu1}
\end{figure}

We have shown that, in order to avoid the vacuum stability constraint in Eq.~\eqref{eq:vac}, $|m_{H_u}^2(M_{\rm GUT})|$ needs to be large. 
This feature is somewhat model independent. 
Then, off-diagonal elements of the slepton mass matrix are induced through the neutrino Yukawa interactions in Eq.\,\eqref{eq:ynu}, which are estimated as~\cite{Hisano:1995nq} (see also appendix \ref{sec:app_a})
\begin{eqnarray}
 (m_{\tilde L}^2)_{ij} \approx \frac{(m_{H_u}^2 + A_u^2)}{8\pi^2}(Y_\nu^\dag)_{ik} \ln (M_{N_k}/M_{\rm GUT}) (Y_\nu)_{kj}.
\label{eq:LFV}
\end{eqnarray}
As for the diagonal element, $(m_{\tilde L}^2)_{33}$ is given by the sum of Eq.~(\ref{eq:mL3}) and (\ref{eq:LFV}) with $i=j=3$.\footnote{The equation \eqref{eq:LFV} is also valid for $i=j$. However, it is subdominant.} 
The neutrino Yukawa coupling $Y_\nu$ is  parameterized as~\cite{Casas:2001sr} 
\begin{eqnarray}
 Y_\nu \left<H_u\right> &=& {\rm diag}(\sqrt{M_{N_1}},\sqrt{M_{N_2}},\sqrt{M_{N_3}})
\, R \, {\rm diag}(\sqrt{m_{\nu_1}},\sqrt{m_{\nu_2}},\sqrt{m_{\nu_3}}) \nonumber \\ 
&\times& {\rm diag}(e^{-i\alpha_1/2},e^{-i\alpha_2/2},1)\,
V_{\rm PMNS}^\dag, 
\end{eqnarray}
where $R$ is a complex orthogonal matrix, $\alpha_1$ and $\alpha_2$ are Majorana phases and $V_{\rm PMNS}$ is the PMNS matrix. 
In the following numerical calculation, we take $R$ to be a real orthogonal matrix, $\alpha_1=\alpha_2=0$ for simplicity. The neutrino mass differences, the mixing angles and the Dirac phase are taken from PDG~\cite{pdg}.
The neutrino parameter dependence of LFV appears in the combination of $\sum_{k}$$(Y_\nu^\dag)_{ik}$$ \ln (M_{N_k}/M_{\rm GUT}) $$(Y_\nu)_{kj}$. 
In figure \ref{fig:nu1}, we check the $m_{\nu_1}$ dependence of the LFV coupling parameter,
$|\sum_{k}$$(Y_\nu^\dag)_{1k}$$ \ln (M_{N_k}/M_{\rm GUT}) $$(Y_\nu)_{k2}|$,
by taking $R$ randomly. Here we take $\tan\beta=20$ and $M_{N_3}=10^{10}\,\mbox{GeV}$ and we consider two cases: i) $M_{N_1}:M_{N_2}:M_{N_3}=1:2:3$ and ii) $M_{N_1}:M_{N_2}:M_{N_3}=1:10:100$. The black line corresponds to the case for $R=1$. We here impose $\sum_{i=1}^3 m_{\nu_i}<0.12\,\mbox{eV}$ \cite{pdg}. We observe that the larger $m_{\nu_1}$ induces the larger LFV. Hereafter we take $m_{\nu_1}=0$ for a conservative estimate.

\begin{figure}[!t]
\begin{center}  
\includegraphics[width=78mm]{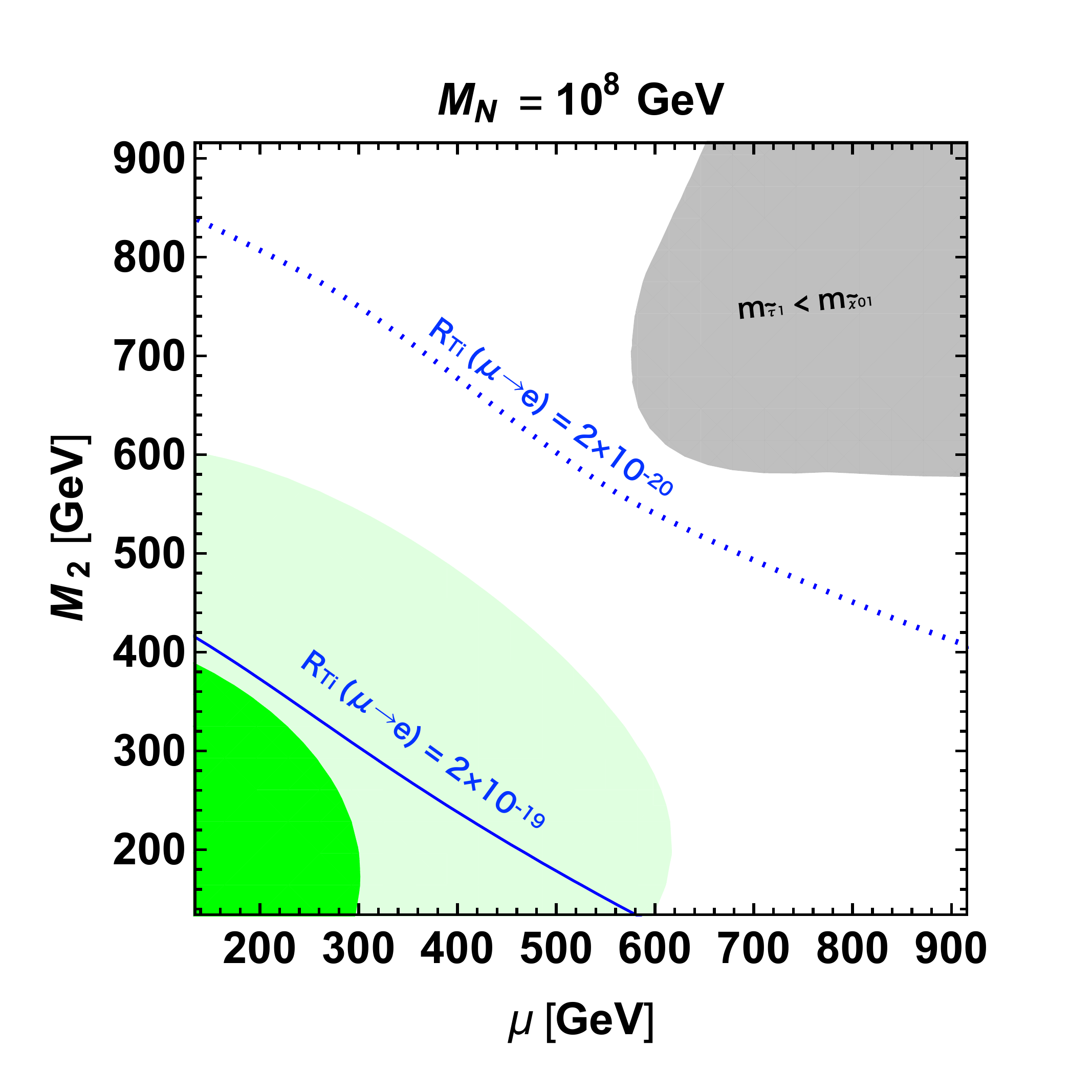}
\includegraphics[width=78mm]{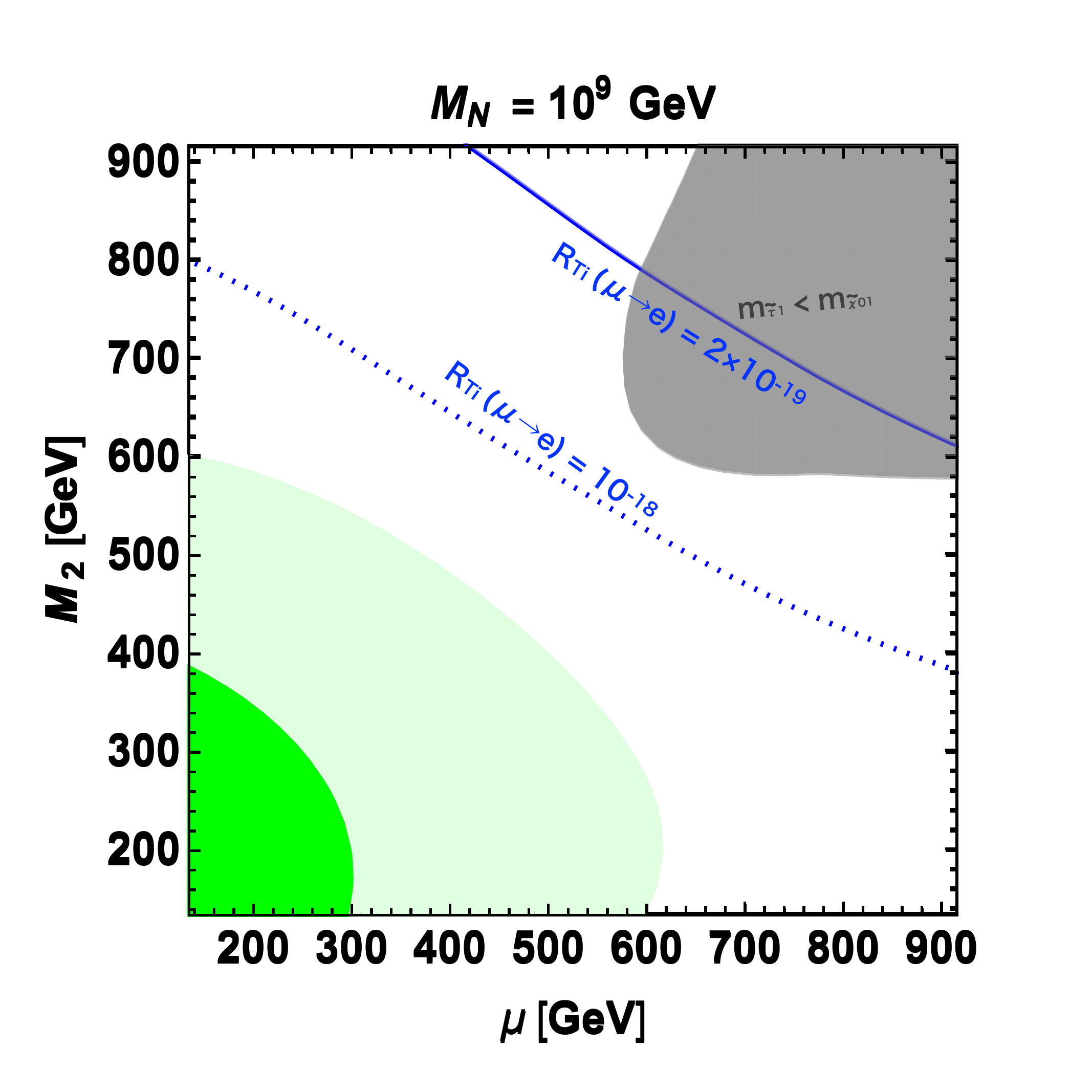}
\includegraphics[width=78mm]{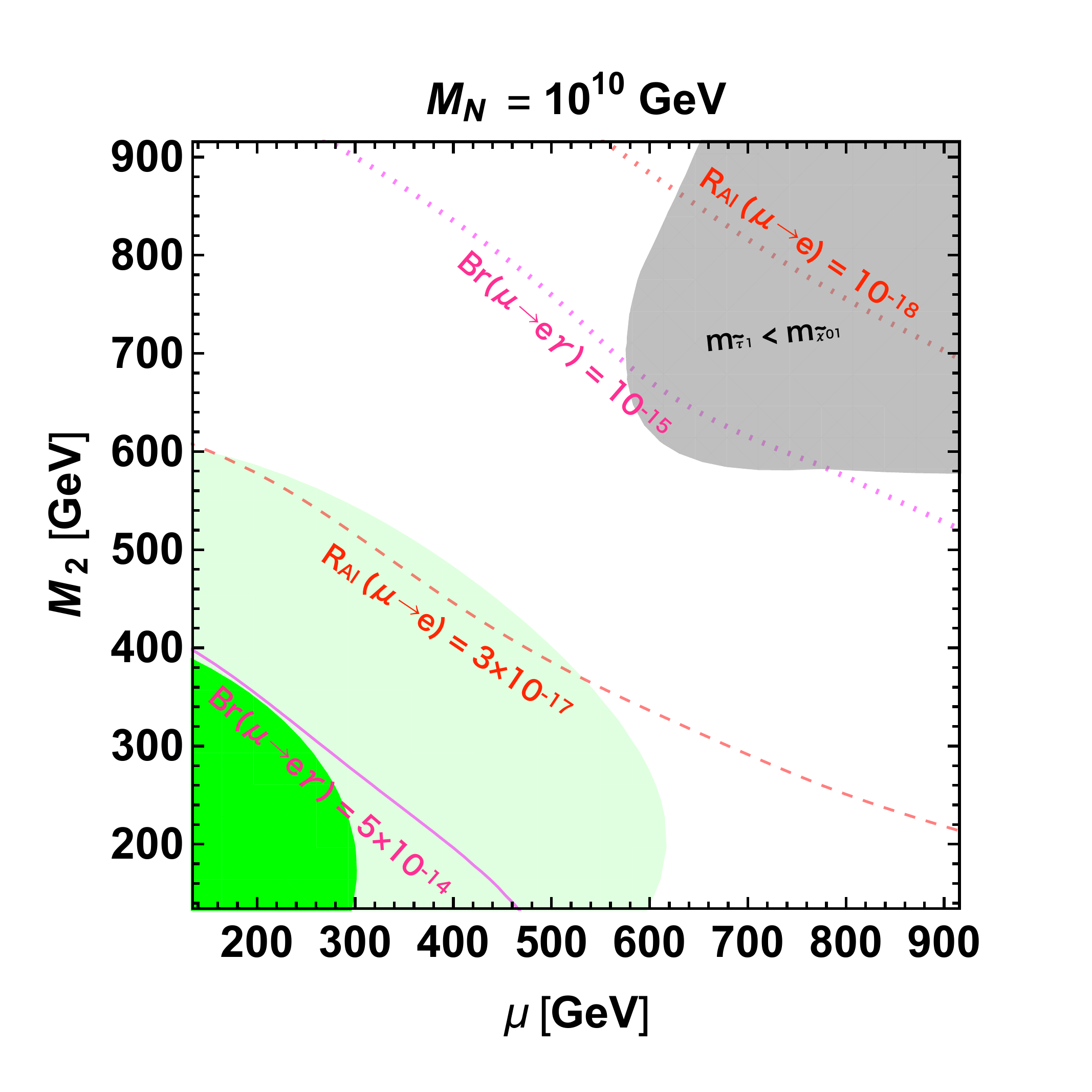}
\includegraphics[width=78mm]{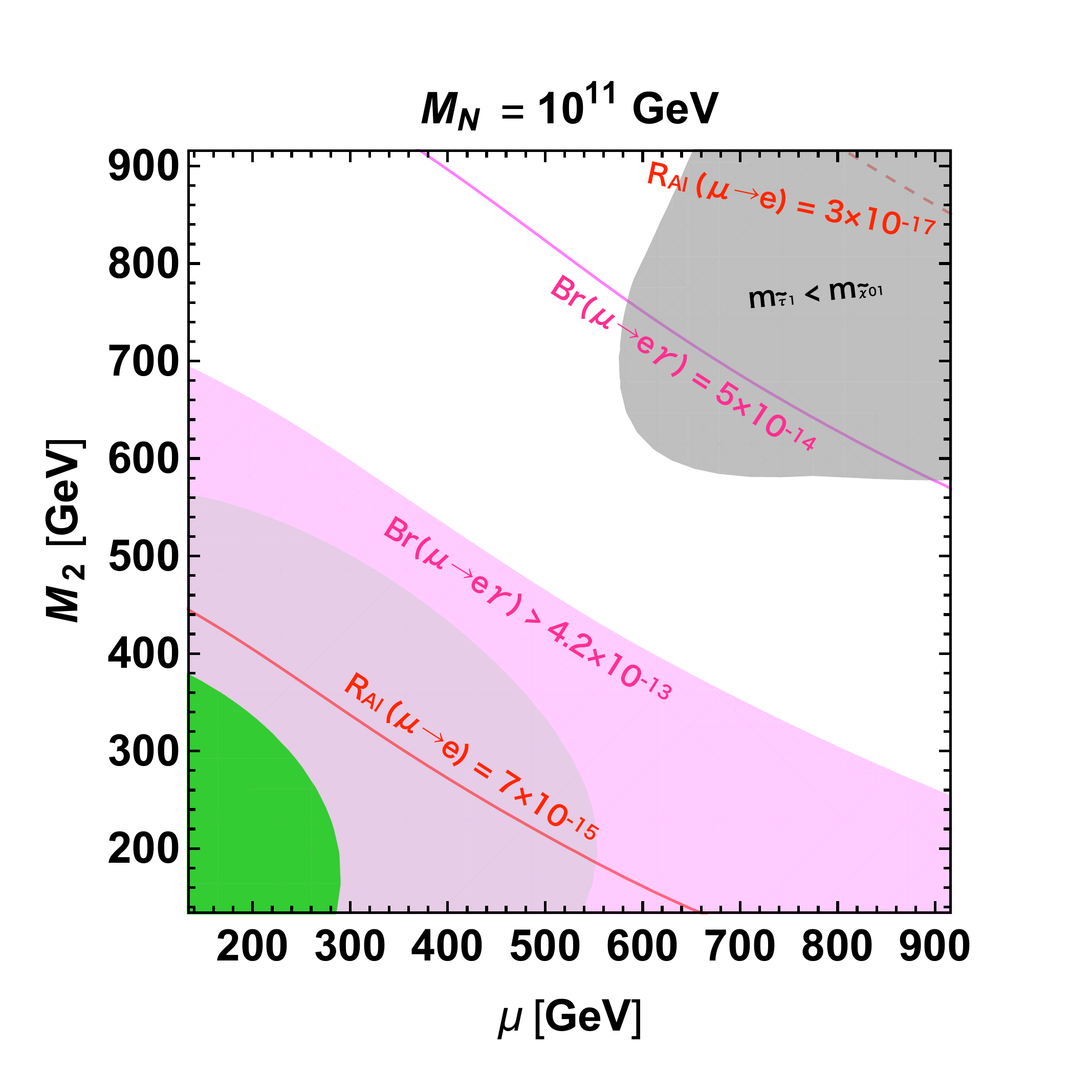}
\end{center}
\caption{Contours of the muon $g-2$ and LFVs in the model {\bf A} with the degenerated RH neutrinos. We take $M_1=3\,\mbox{TeV}$, $M_3=2.5\,\mbox{TeV}$, $m_A=3.4\,\mbox{TeV}$, $A_u=-1\,\mbox{TeV}$ and $\tan\beta=20$.
In the dark (light) green regions, the muon $g-2$ is explained at 1$\sigma$ level (2$\sigma$ level).
The purple shaded regions are excluded due to too large ${\rm Br}(\mu \to e \gamma)$.
The stau becomes the LSP in the gray shaded region.
}
\label{fig:gm2LFV-A}
\end{figure}

\begin{figure}[!t]
\begin{center}  
\includegraphics[width=78mm]{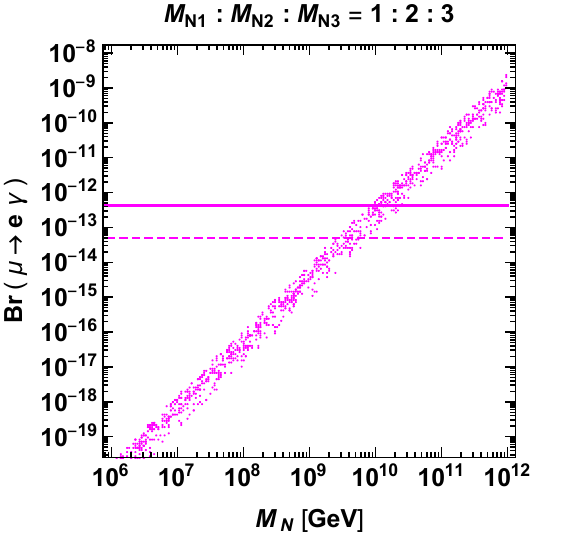}
\includegraphics[width=78mm]{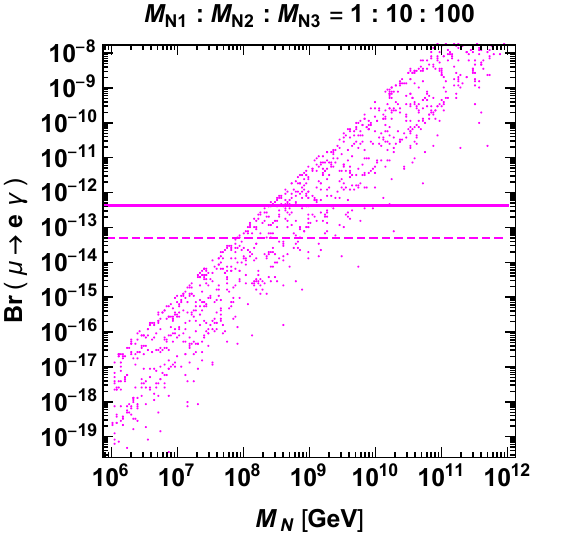}
\end{center}
\caption{${\rm BR}(\mu \to e \gamma)$ is shown for the model {\bf A}. In the left (right) panel, $M_{N_1}:M_{N_2}:M_{N_3}=1:2:3$ 
($M_{N_1}:M_{N_2}:M_{N_3}=1:10:100$) with $M_N=M_{N_1}$.
We fix $M_2=250\,\mbox{GeV}$ and $\mu=260\,\mbox{GeV}$ and the other parameters are same as  figure~\ref{fig:gm2LFV-A}.
}
\label{fig:small_mu_rand1}
\end{figure}

\begin{figure}[!t]
\begin{center}  
\includegraphics[width=78mm]{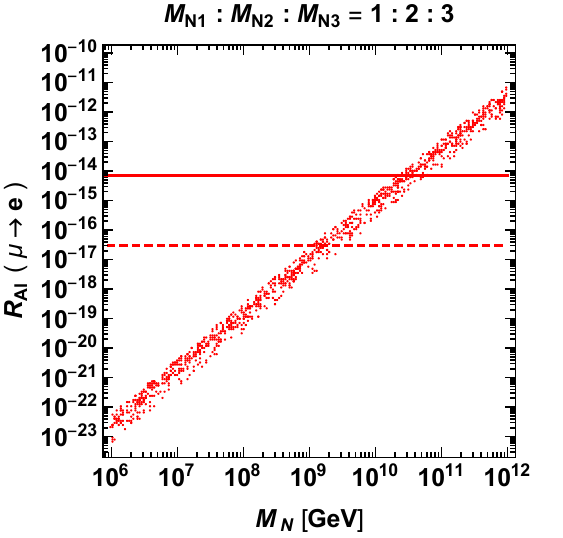}
\includegraphics[width=78mm]{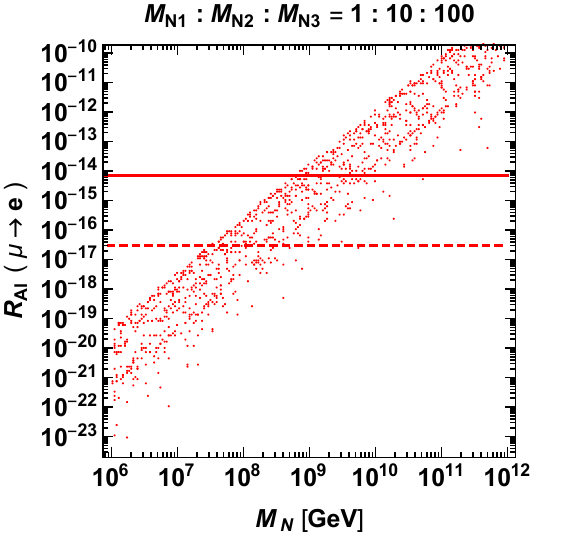}
\end{center}
\caption{${\rm R}_{\rm Al}(\mu \to e)$ is shown for model A.
The parameters are same as figure~\ref{fig:small_mu_rand1}.}
\label{fig:small_mu_rand2}
\end{figure}

\begin{figure}[!t]
\begin{center}  
\includegraphics[width=78mm]{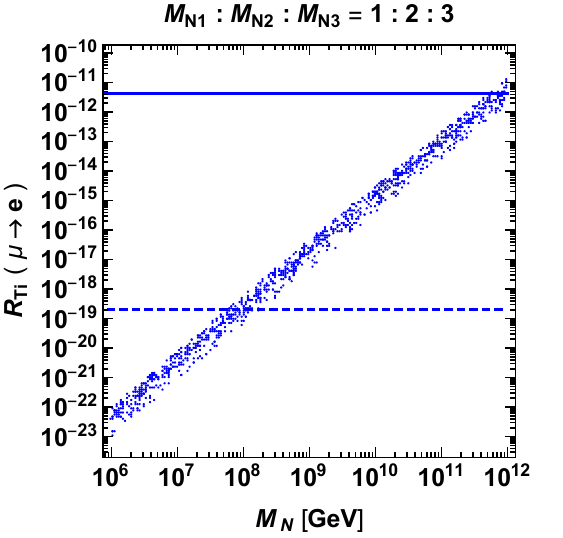}
\includegraphics[width=78mm]{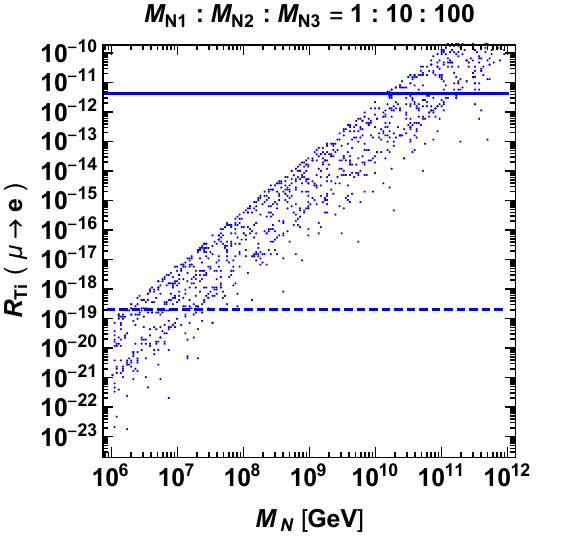}
\end{center}
\caption{${\rm R}_{\rm Ti}(\mu \to e)$ is shown for the model {\bf A}.
The parameters are same as figure~\ref{fig:small_mu_rand1}.}
\label{fig:small_mu_rand3}
\end{figure}

\begin{table*}[!t]
\caption{A mass spectrum in the model {\bf A}. We take $M_{N_1}=M_{N_2}=M_{N_3}=10^9$\,GeV.}
\label{tab:1}
\begin{center}
\begin{tabular}{|c||c|}
\hline
Parameters & Point {\bf I} \\ 
\hline
$M_1$\,(GeV) & 3000 \\
$M_2$\,(GeV) & 250 \\
$M_3$\,(GeV) & 2500 \\
$A_u$\,(GeV) & -1000 \\
$\mu$\,(GeV) & 260 \\
$m_A$\,(GeV) & 3400 \\
$\tan\beta$  & 20  \\
\hline
Particles & Mass (GeV) \\
\hline
$\tilde{g}$ & 5090  \\
$\tilde{q}$  & 4330-4400 \\
$\tilde{t}_{1,2}$ & 3380, 3840 \\
$\tilde{b}_{1,2}$  & 3850, 4300 \\
$\tilde{e}_{L, R}$ & 500, 1110 \\
$\tilde{\mu}_{L,R}$ &499, 1110 \\
$\tilde{\tau}_{1,2}$ & 229, 908 \\
$\tilde{\chi}_{1,2,3}^0$ & 144, 273, 293 \\
$\tilde{\chi}_{4}^0$ & 1340 \\
$\tilde{\chi}^{\pm}_{1,2}$ & 145, 299 \\
$h_{\rm SM\mathchar`-like}$ & 125.1 \\
\hline
$10^{9}\Delta a_\mu $& 2.19 \\
$m_{H_u}^2 (M_{\rm GUT})$(GeV$^2$) & $1.17\times 10^7$\\
$m_{H_d}^2 (M_{\rm GUT})$(GeV$^2$) & $1.23\times 10^7$\\
\hline
\end{tabular}
\end{center}
\end{table*}

\begin{figure}[!t]
\begin{center}  
\includegraphics[width=78mm]{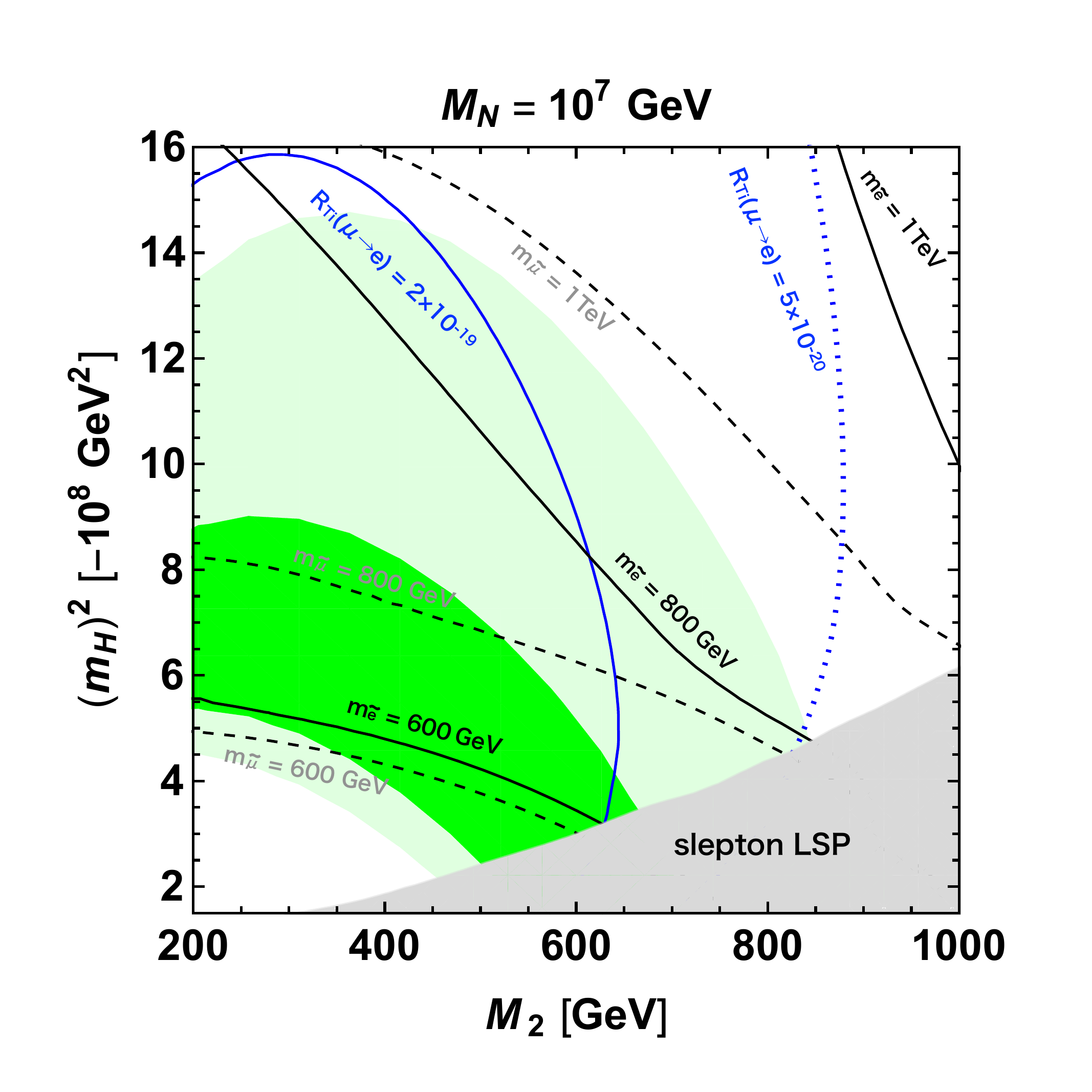}
\includegraphics[width=78mm]{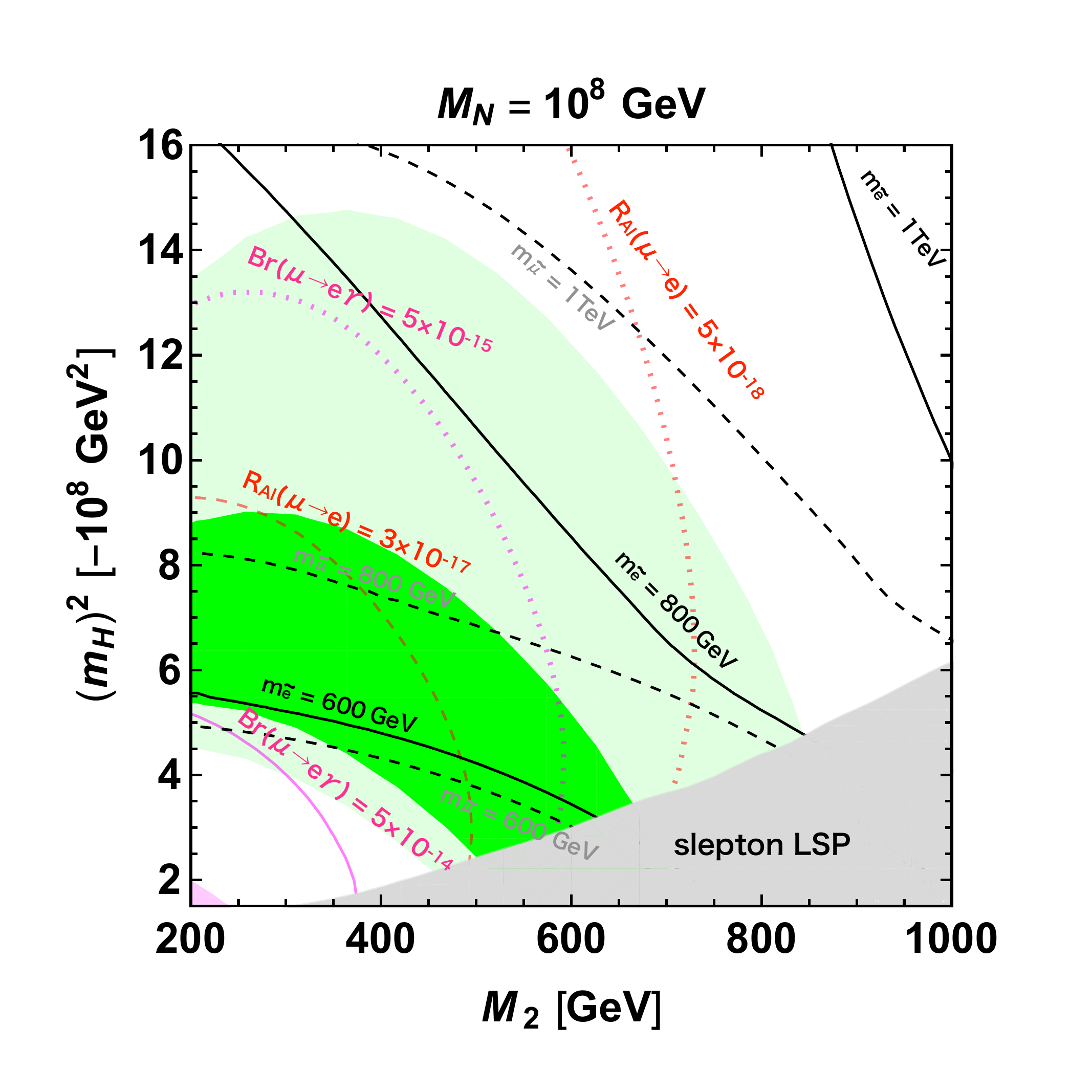}
\includegraphics[width=78mm]{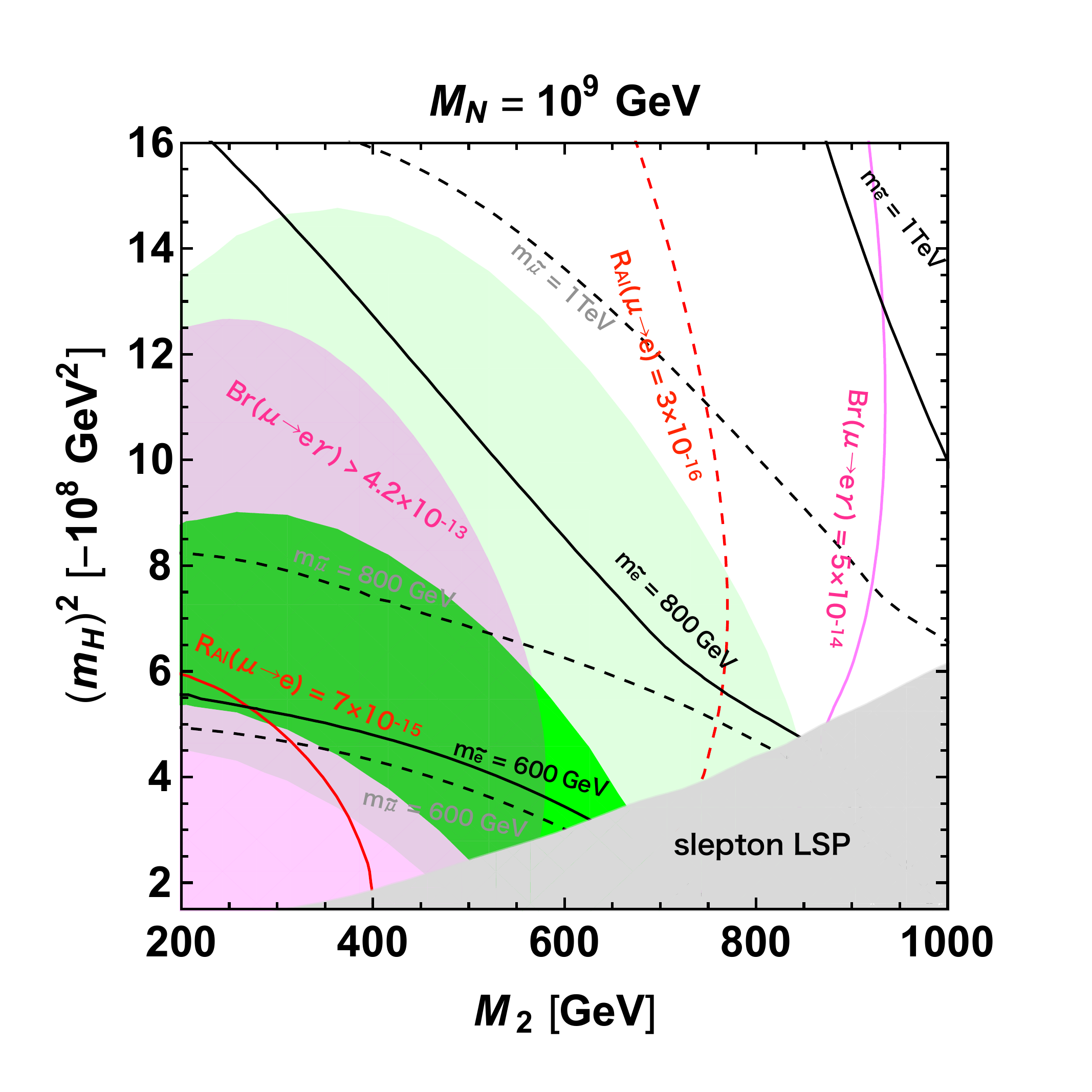}
\includegraphics[width=78mm]{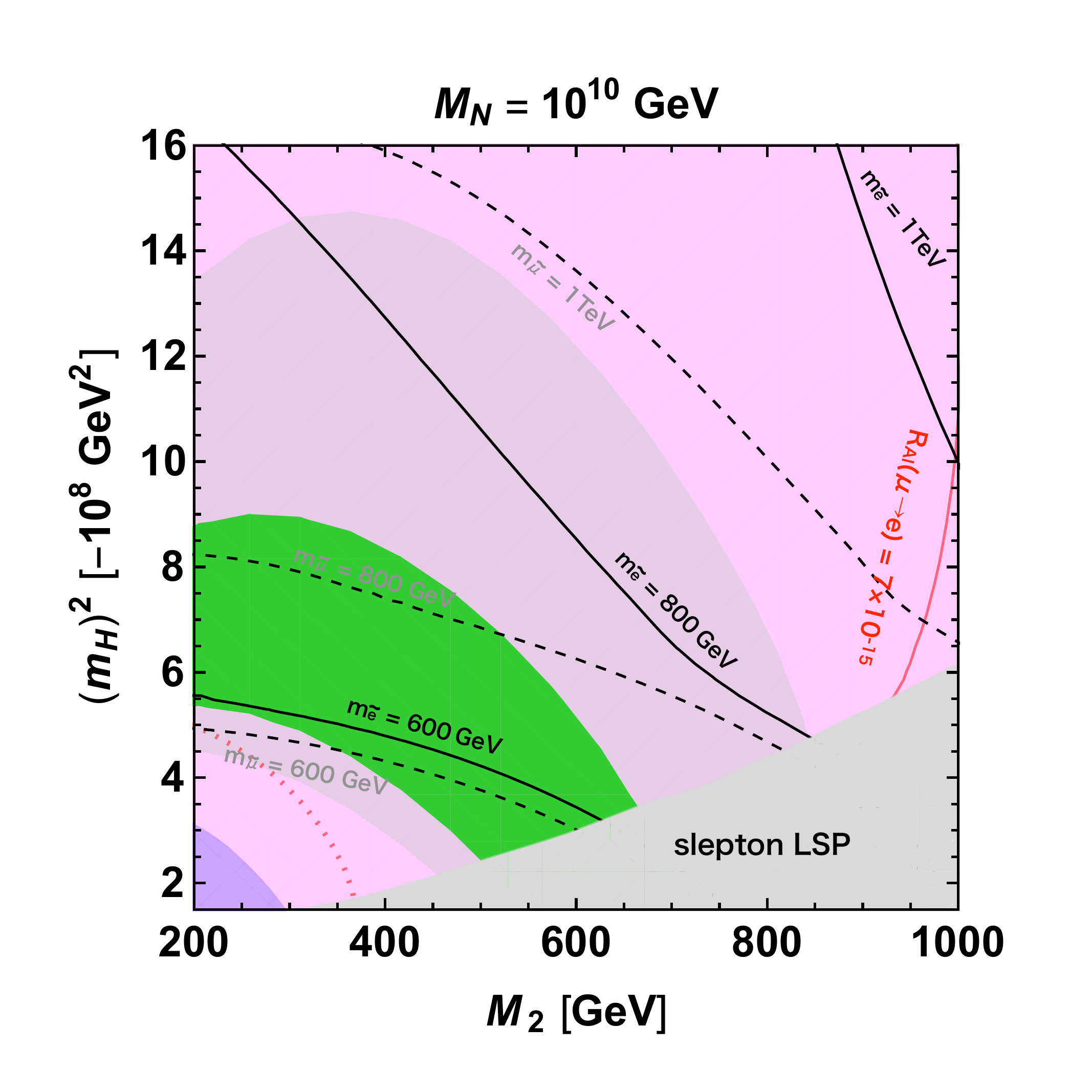}
\end{center}
\caption{Contours of the muon $g-2$ and LFVs in the model {\bf B} with the degenerated RH neutrinos. We take $M_3=-4\,\mbox{TeV}$, $\tan\beta=40$ and $m^2_{H_d}=m^2_{H_u}(=m^2_{H})$.}
\label{fig:gm2LFV-B}
\end{figure}

\begin{figure}[!t]
\begin{center}  
\includegraphics[width=78mm]{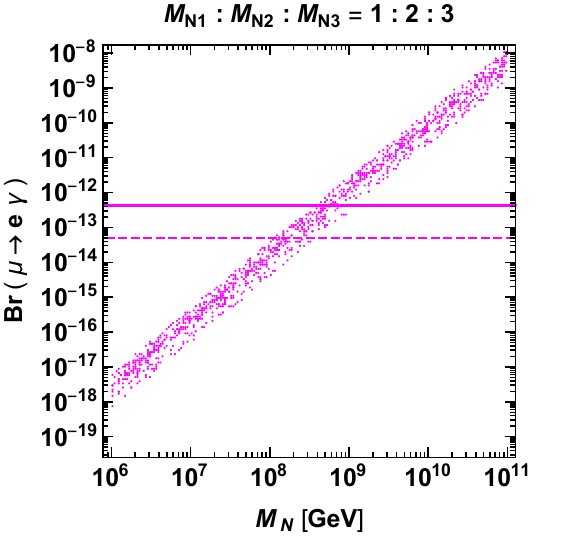}
\includegraphics[width=78mm]{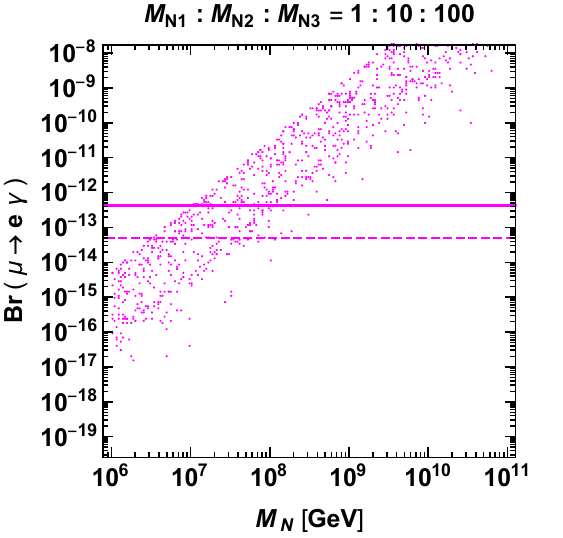}
\end{center}
\caption{${\rm BR}(\mu \to e \gamma)$ is shown for the model {\bf B}. In the left (right) panel, $M_{N_1}:M_{N_2}:M_{N_3}=1:2:3$ 
($M_{N_1}:M_{N_2}:M_{N_3}=1:10:100$) with $M_N=M_{N_1}$.
We take $m^2_{H_u}=m^2_{H_d}=-4\times 10^8\,\mbox{GeV}^2$ and $M_2=600\,\mbox{GeV}$. The other parameters are same as figure~\ref{fig:gm2LFV-B}.}
\label{fig:largemu_binowino_rand1}
\end{figure}

\begin{figure}[!t]
\begin{center}  
\includegraphics[width=78mm]{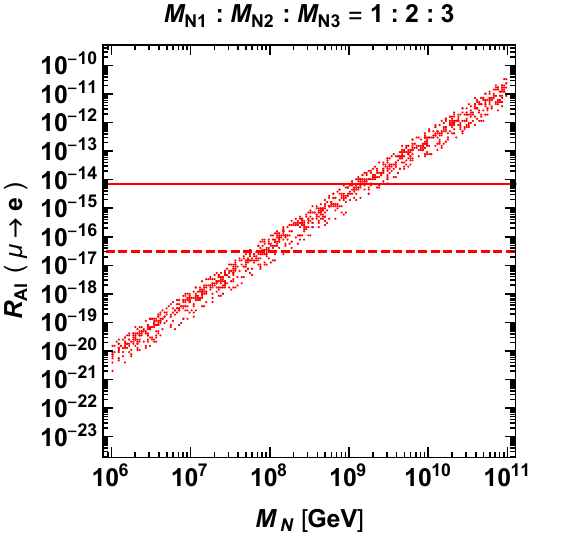}
\includegraphics[width=78mm]{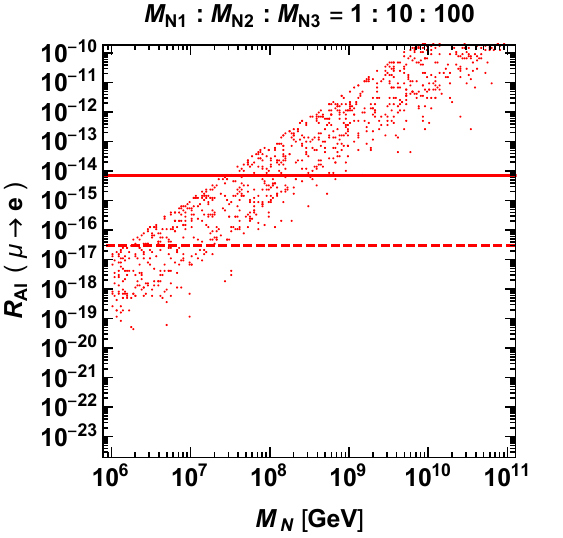}
\end{center}
\caption{${\rm R}_{\rm Al}(\mu \to e)$ is shown for the model {\bf B}.
The parameters are same as figure~\ref{fig:largemu_binowino_rand1}.}
\label{fig:largemu_binowino_rand2}
\end{figure}

\begin{figure}[!t]
\begin{center}  
\includegraphics[width=78mm]{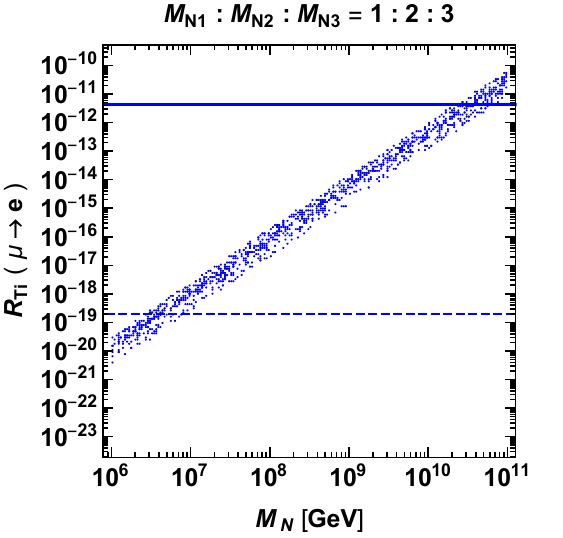}
\includegraphics[width=78mm]{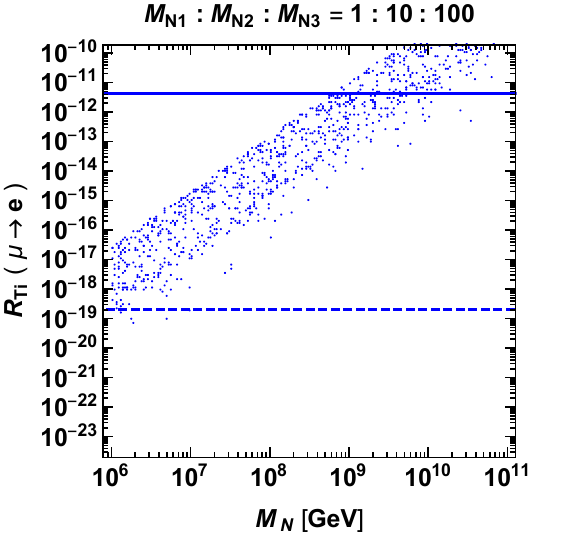}
\end{center}
\caption{${\rm R}_{\rm Ti}(\mu \to e)$ is shown for the model {\bf B}.
The parameters are same as figure~\ref{fig:largemu_binowino_rand1}.}
\label{fig:largemu_binowino_rand3}
\end{figure}

\begin{table*}[!t]
\caption{Mass spectra in the model {\bf B} and {\bf C}. We take $M_{N_1}=M_{N_2}=M_{N_3}=10^9$\,GeV. }
\label{tab:2}
\begin{center}
\begin{tabular}{|c||c|c|}
\hline
Parameters & Point {\bf II} & Point {\bf III} \\ 
\hline
$M_2$ (GeV) & 600  & 2000  \\
$M_3$ (GeV) & $-4000$  & $-4000$  \\
$m_{H_u}^2$ (GeV$^2$) & $-4 \times 10^8$ & $-10^8$\\
$\tan\beta$  & 40  & 40 \\
\hline
Particles & Mass (GeV) & Mass (GeV) \\
\hline
\hline
$\tilde{g}$ & 8150 &  8040\\
$\tilde{q}$  & 6650-6670 & 6700-6810\\
$\tilde{t}_{1,2}$ (TeV) & 10.4, 10.6 & 7.4, 7.5\\
$\tilde{b}_{1,2}$ (TeV) & 10.6, 11.0 & 7.3, 7.5\\
$\tilde{e}_{L, R}$ & 632, 675 & 1260, 197\\
$\tilde{\mu}_{L,R}$ & 661, 728 & 1270, 268 \\
$\tilde{\tau}_{1,2}$ (TeV) & 4.5, 6.4 & 2.7, 3.4\\
$\tilde{\chi}_{1,2}^0$ & 577, 605 & 190, 1790\\
$\mu$ (TeV) & 17.4 & 9.2\\
$\tilde{\chi}^{\pm}_{1}$ & 605 & 1790\\
$h_{\rm SM\mathchar`-like}$ & 125.1 & 125.2\\
$H_{A}$ (TeV) & 5.8 & 2.1 \\
\hline
$10^{9}\Delta a_\mu $& 2.26 & 2.20 \\
$\Omega_{\rm DM} h^2$ & 0.119   &  0.120   \\
\hline
\end{tabular}
\end{center}
\end{table*}

\begin{figure}[!t]
\begin{center}  
\includegraphics[width=78mm]{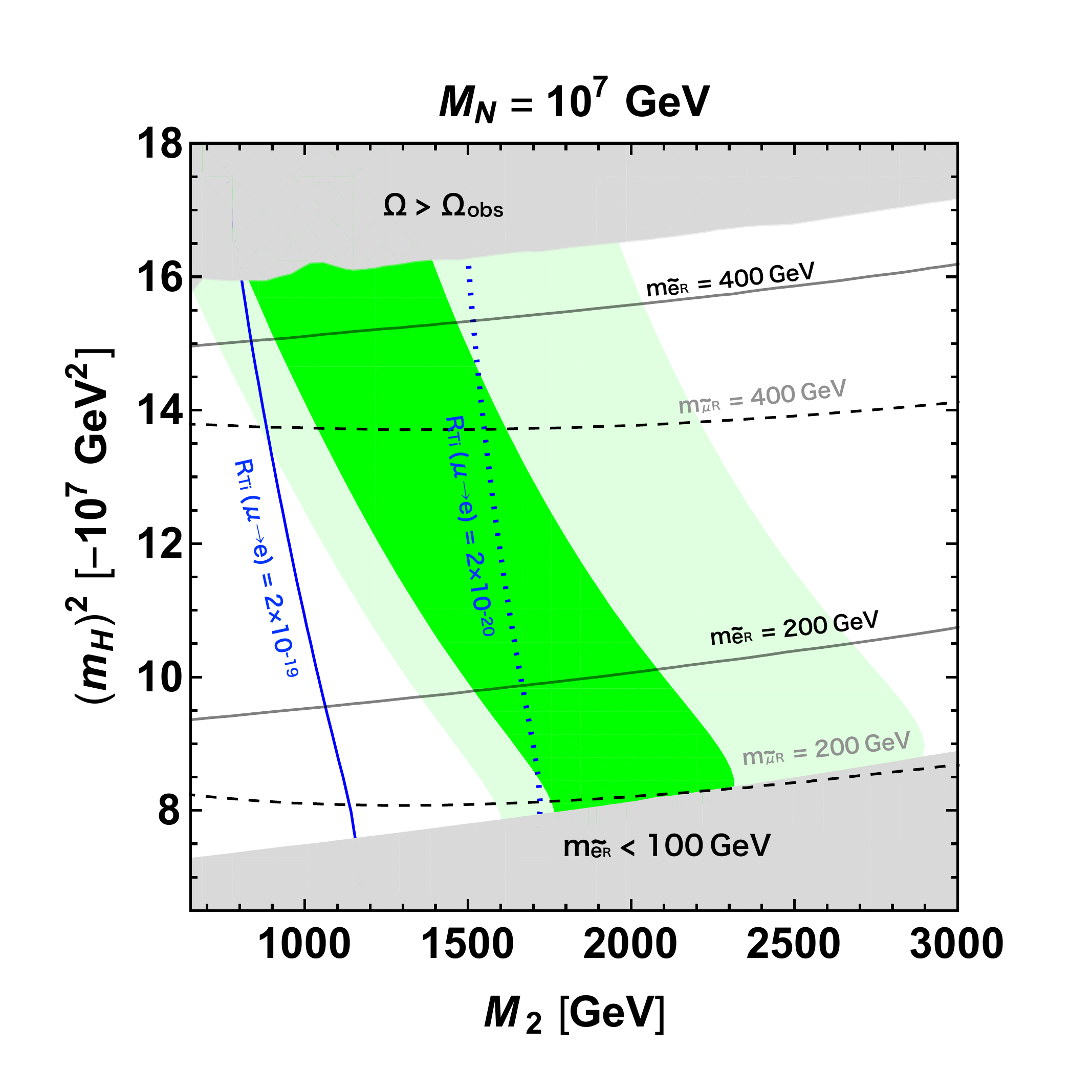}
\includegraphics[width=78mm]{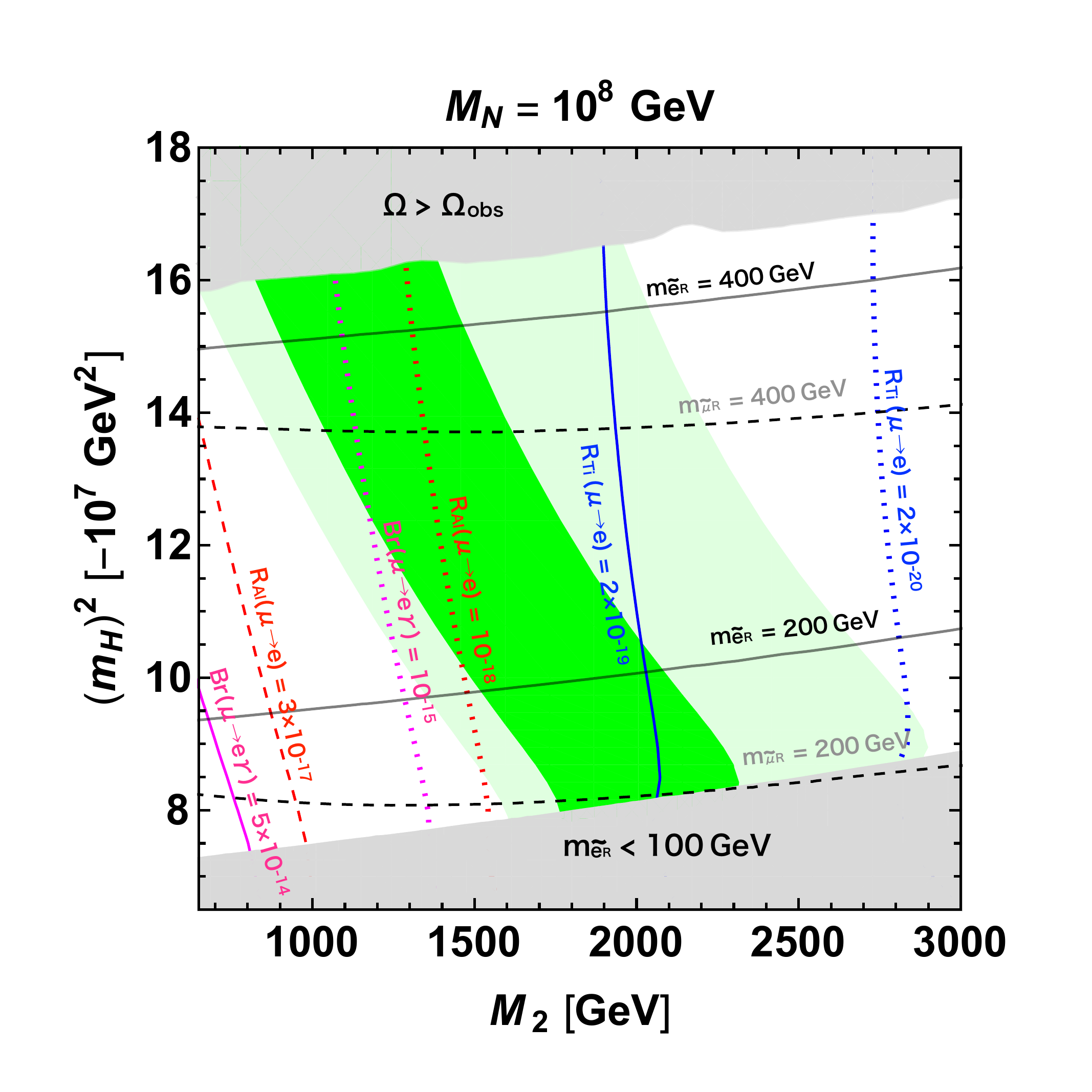}
\includegraphics[width=78mm]{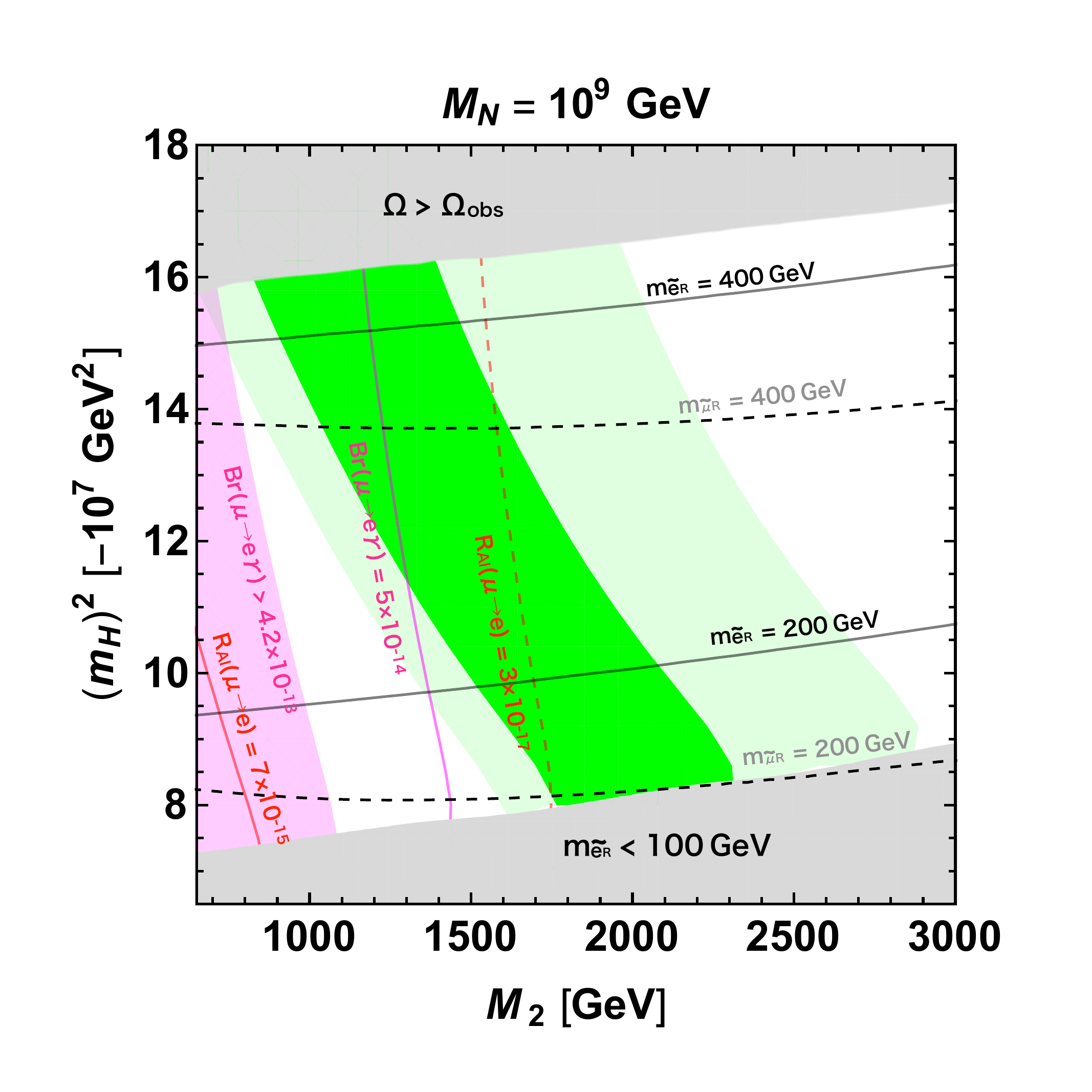}
\includegraphics[width=78mm]{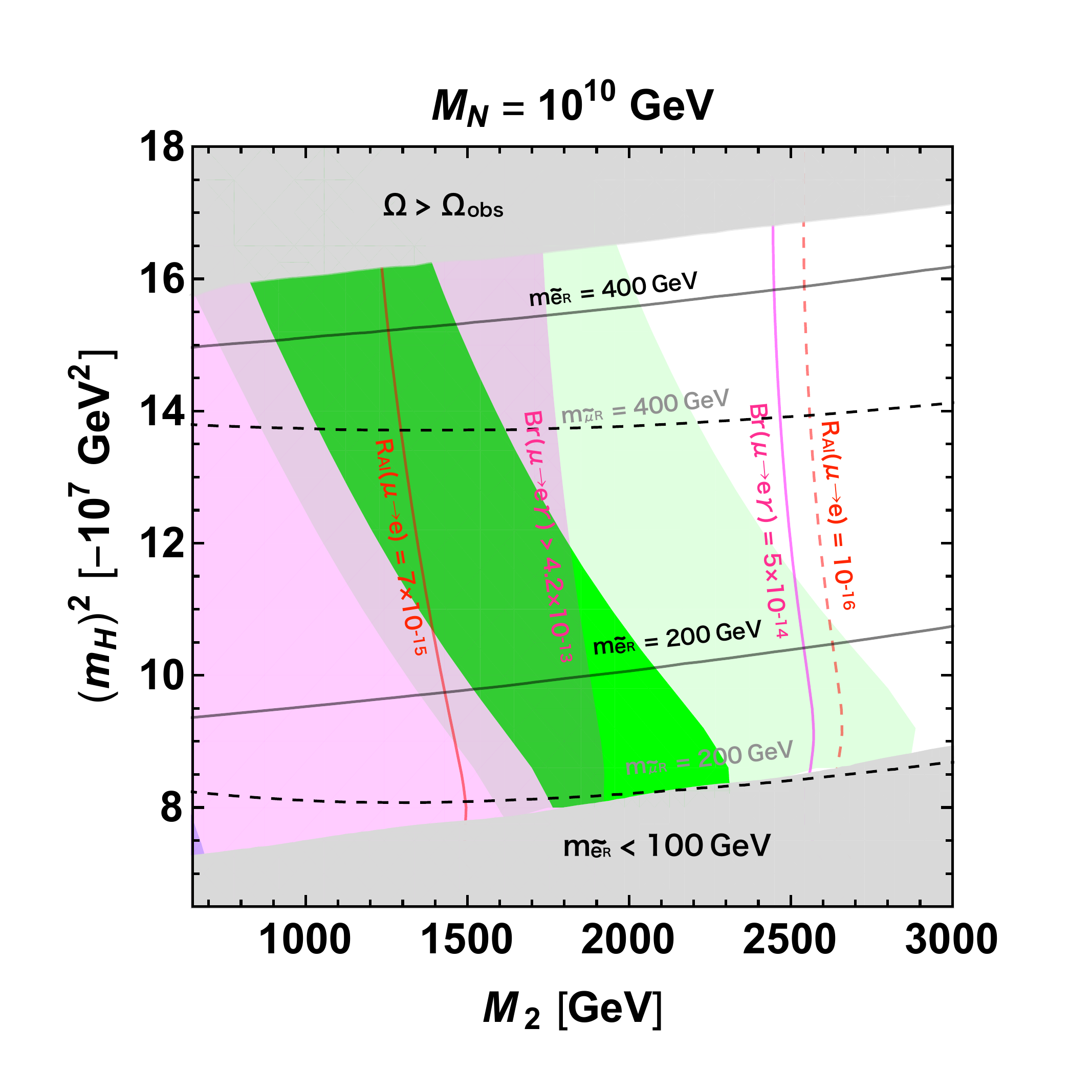}
\end{center}
\caption{Contours of the muon $g-2$ and LFVs in the model {\bf C} with the degenerated RH neutrinos. The parameters are same as figure~\ref{fig:gm2LFV-B}.}
\label{fig:gm2LFV-C}
\end{figure}

\begin{figure}[!t]
\begin{center}  
\includegraphics[width=78mm]{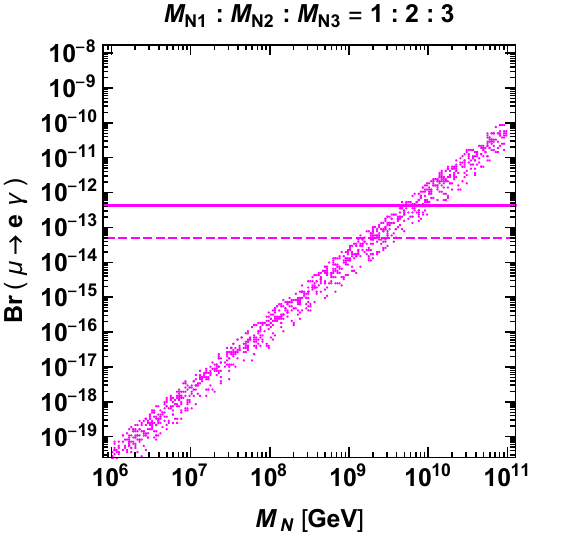}
\includegraphics[width=78mm]{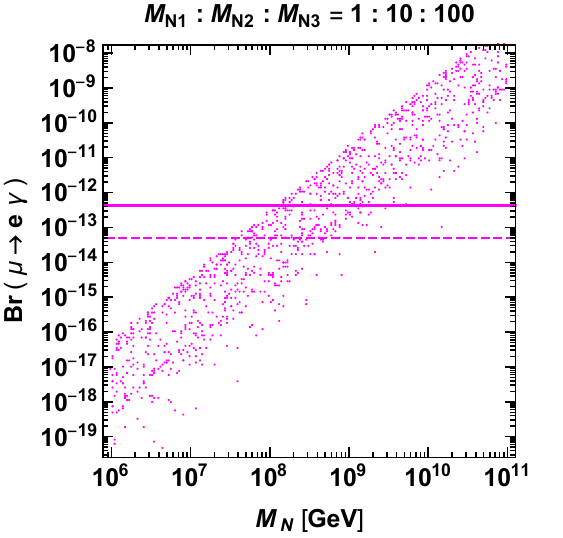}
\end{center}
\caption{${\rm BR}(\mu \to e \gamma)$ is shown for the model {\bf C}. In the left (right) panel, $M_{N_1}:M_{N_2}:M_{N_3}=1:2:3$ 
($M_{N_1}:M_{N_2}:M_{N_3}=1:10:100$) with $M_N=M_{N_1}$.
We take $m^2_{H_u}=m^2_{H_d}=-10^8\,\mbox{GeV}^2$ and $M_2=2\,\mbox{TeV}$. The other parameters are taken as same as figure \ref{fig:gm2LFV-C}
}
\label{fig:largemu_binoslep_rand1}
\end{figure}

\begin{figure}[!t]
\begin{center}  
\includegraphics[width=78mm]{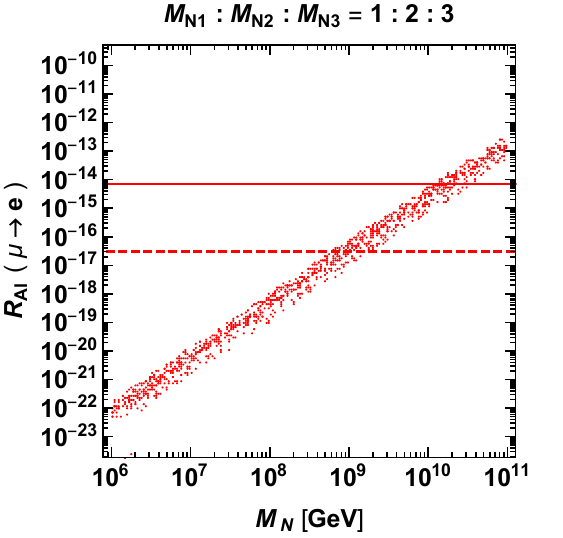}
\includegraphics[width=78mm]{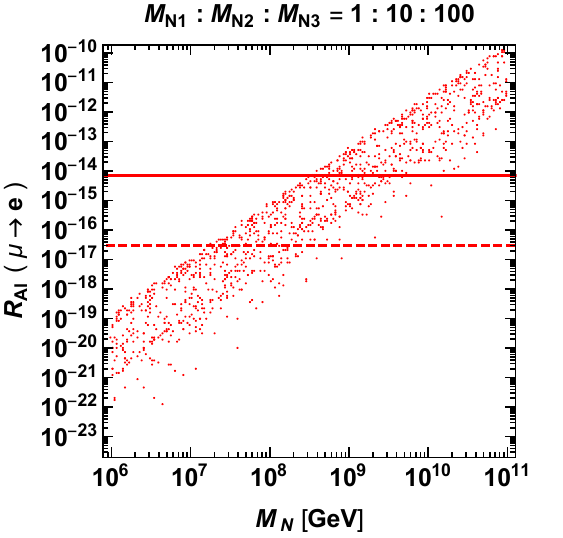}
\end{center}
\caption{${\rm R}_{\rm Al}(\mu \to e)$ is shown for the model {\bf C}.
The parameters are same as figure~\ref{fig:largemu_binoslep_rand1}.}
\label{fig:largemu_binoslep_rand2}
\end{figure}

\begin{figure}[!t]
\begin{center}  
\includegraphics[width=78mm]{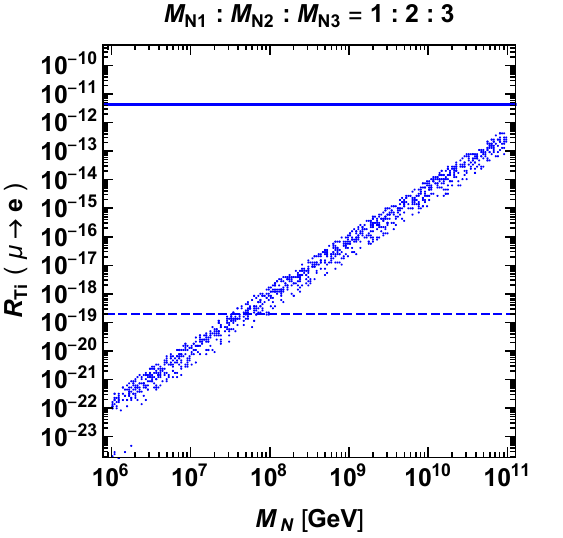}
\includegraphics[width=78mm]{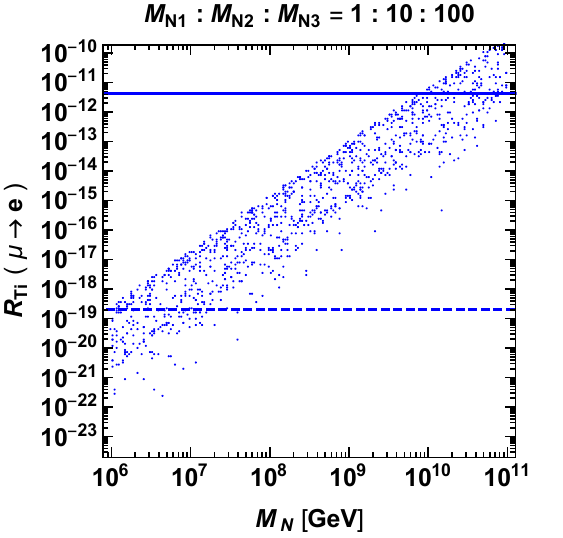}
\end{center}
\caption{${\rm R}_{\rm Ti}(\mu \to e)$ is shown for the model {\bf C}.
The parameters are same as figure~\ref{fig:largemu_binoslep_rand1}.}
\label{fig:largemu_binoslep_rand3}
\end{figure}

Once we obtain the non-negligible off-diagonal elements of the slepton mass matrix, the LFV processes, such as $\mu \to e \gamma$ and $\mu \to e$ conversion, are induced \cite{Hisano:1995cp}. 
In what follows, we estimate the sizes of the LFV processes for each models, and discuss impacts on future experiments. 
We employ {\tt{SuSpect 2.43}}~\cite{Djouadi:2002ze} to calculate the spectrum for SUSY particles. Combining the output of {\tt{SuSpect 2.43}} and the general formulae given in Ref.~\cite{Hisano:1995cp}, we estimate the sizes of the LFVs. The SM-like Higgs mass and the DM relic density are estimated using {\tt{FeynHiggs 2.14.3}}~\cite{Heinemeyer:1998yj, Heinemeyer:1998np, Degrassi:2002fi, Frank:2006yh, Hahn:2013ria, Bahl:2016brp, Bahl:2017aev, Bahl:2018qog,Bahl:2019ago} and {\tt{MicrOmegas 5.0.4}}~\cite{Belanger:2001fz,Belanger:2004yn}, respectively. 
For the estimation of muon $g-2$, we include the dominant two-loop corrections: the logarithmic QED correction~\cite{Degrassi:1998es} and the $\tan\beta$ enhanced correction to the muon Yukawa coupling~\cite{Marchetti:2008hw}. The two-loop corrections can be large as $\mathcal{O}(10)\%$.

Let us first focus on model {\bf{A}}, where the vacuum stability constraint is avoided with the small $\mu$. 
Figure \ref{fig:gm2LFV-A} shows the sizes of the muon $g-2$ and LFV processes for $M_{N_1}=M_{N_2}=M_{N_3}(=M_{N})$. 
In this case, the LFV processes become independent of $R$ ($Y_\nu^\dag Y_\nu$ is independent of $R$). As for the SUSY breaking parameters and $\tan\beta$, we take $M_1=3\,\mbox{TeV}$, $M_3=2.5\,\mbox{TeV}$, $m_A=3.4\,\mbox{TeV}$, $A_u=-1\,\mbox{TeV}$ and $\tan\beta=20$. We fix the RH neutrino mass as $10^8\,\mbox{GeV}$, $10^9\,\mbox{GeV}$, $10^{10}\,\mbox{GeV}$, and $10^{11}\,\mbox{GeV}$ in the top-left, top-right, bottom-left, and bottom-right figures, respectively. 
In the gray shaded region, there should be severe constraints from LHC because the stau becomes the lightest SUSY particle (LSP) and long-lived~\cite{Kutzner:2639964,Aaboud:2019trc}.
The muon $g-2$ is explained at $1\sigma$ level ($2\sigma$ level) in the dark (light) green region. We see that the model {\bf{A}} can explain muon $g-2$ if we take $M_2, \mu\simeq \mathcal{O}(100\,\mbox{GeV})$. On the other hand, our model simultaneously predicts sizable LFV processes as we discussed above. The current limit on the LFV processes is shown by the purple shaded region, which is given by MEG experiment \cite{TheMEG:2016wtm}. We observe that the MEG experiment excludes the parameter region for muon $g-2$ when $M_{N}=10^{11}\,\mbox{GeV}$. Future sensitivities on the relevant LFV processes, on the other hand, are shown by the colored lines. The region below these lines can be tested by future LFV experiments. The purple line corresponds to the future sensitivity of $\mu\to e\gamma$, $\mbox{Br}(\mu\to e\gamma)\approx 5 \times 10^{-14}$, at MEG-II~\cite{Cattaneo:2017psr}. The red and dashed red lines are the future sensitivities of $\mu -e $ conversion in Al at COMET Phase-I \cite{Adamov:2018vin} and COMET phase II~\cite{comet-tdr}, which correspond to ${R}_{\rm{Al}}(\mu\to e)\approx  7\times 10^{-15}$ and $\approx 3\times 10^{-17}$ respectively. Mu2e~\cite{Miscetti:2020gkk} gives similar sensitivity as COMET phase II. 
The blue line is the future sensitivity of $\mu \to e$ conversion in Ti, $R_{\rm{Ti}}(\mu\to e)\approx 2 \times 10^{-19}$, at PRISM/PRIME~\cite{Kuno:2012pt}. We find that the future LFV experiments can investigate the parameter region for muon $g-2$ if the RH neutrinos are heavier than $10^8\,\mbox{GeV}$.

In table \ref{tab:1}, we show the typical mass spectrum in this model. Here we fix the RH neutrino masses as $M_{N_1}=M_{N_2}=M_{N_3}=10^9\,\mbox{GeV}$. We note, however, that the SUSY mass spectrum is almost insensitive to the masses of the RH neutrinos.

It should be reminded that, if we relax the degeneracy of the RH neutrino masses, the size of LFV depends on the structure of a matrix $R$ which cannot be determined by observables. Let us estimate the $R$ dependence of LFV by taking $R$ randomly. The results are given by figures \ref{fig:small_mu_rand1}, \ref{fig:small_mu_rand2} and \ref{fig:small_mu_rand3} which show the size of $\mbox{Br}(\mu\to e\gamma)$, $R_{\rm{Ti}}(\mu\to e)$ and $R_{\rm{Al}}(\mu\to e)$ as a function of the mass of the lightest RH neutrino, respectively. We consider the two cases; i) the case where the RH neutrinos are almost degenerate, namely $M_{N_1}:M_{N_2}:M_{N_3}=1:2:3$ and ii) the mass spectrum for the RH neutrinos is hierarchical, namely $M_{N_1}:M_{N_2}:M_{N_3}=1:10:100$. In both cases, we fix $M_2=250\,\mbox{GeV}$ and $\mu=260\,\mbox{GeV}$ and the other parameters are same as figure \ref{fig:gm2LFV-A}. We see that the degeneracy of the RH neutrino masses reduces the dependence of $R$. In these figures, we also show the current limits and the future sensitivities on the LFV. The purple and dashed purple lines in figure \ref{fig:small_mu_rand1} are the current limit by MEG \cite{TheMEG:2016wtm} and the future sensitivity at MEG-II~\cite{Cattaneo:2017psr}. The red and dashed red lines in figure \ref{fig:small_mu_rand2} show the future sensitivities of $R_{\rm{Al}}(\mu\to e)$ which are same as figure \ref{fig:gm2LFV-A}. The blue and dashed blue lines in figure \ref{fig:small_mu_rand1} are the current upper limit by SINDRUM II, $R_{\rm{Ti}}(\mu \to e)\approx  4.4\times 10^{-12} $, \cite{Badertscher:1990pu} and the future sensitivity at MEG-II~\cite{Cattaneo:2017psr}.

We next consider the model {\bf{B}}. In this model, $\mu\tan\beta$ is large but the vacuum stability constraint is avoided thanks to the heavy staus. Figure \ref{fig:gm2LFV-B} shows the contours of the muon $g-2$ and LFV in the $(m^2_{H_u}, M_2)$ plane. We also show the contours of the mass of the lighter selectron and smuon, ${m}_{\tilde{e}}$ and ${m}_{\tilde{\mu}}$, as the black and black-dotted lines. Here we take $M_3=-4\,\mbox{TeV}$, $\tan\beta=40$, $m^2_{H_d}=m^2_{H_u}$, $M_{N_1}=M_{N_2}=M_{N_3}$, and $M_1$ is fixed as we obtain the correct DM relic density. We need to avoid the gray shaded region because slepton becomes LSP in the region. 
The color notation for the muon $g-2$ and LFV is same with figure \ref{fig:gm2LFV-A}. The blue shaded region in the bottom-right figure is the current exclusion limit given by SINDRUM II, $R_{\rm{Ti}}(\mu \to e)\approx 4.3\times 10^{-12} $, \cite{Badertscher:1990pu}.  
We see that the model {\bf{B}} can explain muon $g-2$, and the favorable parameter region can be tested by future LFV experiments if the RH neutrino are heavier than $10^7\,\mbox{GeV}$.
The typical mass spectrum in this model is summarized in the left-handed side of the table \ref{tab:2}. 
The $R$ dependence of the LFV in the model {\bf{B}} is shown by figures \ref{fig:largemu_binowino_rand1}, \ref{fig:largemu_binowino_rand2}, and \ref{fig:largemu_binowino_rand3}. We here take $m^2_{H_u}=m^2_{H_d}=-4\times 10^8\,\mbox{GeV}^2$ and $M_2=600\,\mbox{GeV}$. The other parameters are taken as same as \ref{fig:gm2LFV-B}.

Finally, we discuss the model {\bf{C}}. In this model, the vacuum stability constraint is also avoided thanks to the heavy staus. 
In figure \ref{fig:gm2LFV-C}, we take the same parameters as figure \ref{fig:gm2LFV-B}, but we focus on the different $(m^2_{H_u}, M_2)$ region where the correct DM relic density can be realized by bino-slepton coannihilation process. The gray shaded region should be avoided because selectron becomes lighter than $100\,\mbox{GeV}$ in the lower gray shaded region, while the DM abundance becomes larger than the observed value in the upper gray shaded region. We observe that the model {\bf{C}} can explain muon $g-2$ keeping the consistency with the current LFV measurements, and the future LFV experiments can test the parameter region where $M_{N_1}>10^8\,\mbox{GeV}$. 
We show the typical mass spectrum of this model in the right-handed side of the table \ref{tab:2}. 
Figures \ref{fig:largemu_binoslep_rand1}, \ref{fig:largemu_binoslep_rand2}, and \ref{fig:largemu_binoslep_rand3} show the $R$ dependence of the LFV in the model {\bf{C}}. We here take $m^2_{H_u}=m^2_{H_d}=-10^8\,\mbox{GeV}^2$ and $M_2=2\,\mbox{TeV}$. The other parameters are taken as same as figure \ref{fig:gm2LFV-C}.

\section{Conclusion}
In this paper, we have shown that, in SUSY models explaining the muon $g-2$ anomaly, $\mu \to e \gamma$ and $\mu \to e$ conversion are very likely to be observed at the future experiments if the RH neutrinos are heavier than $10^9$\,GeV, motivated by the successful thermal leptogenesis. We have observed $\mbox{BR}(\mu\to e\gamma)\gtrsim 10^{-14}-10^{-12}$, $\mbox{R}_{\rm{Al}}(\mu\to e)\gtrsim 10^{-17}-10^{-14}$, and $\mbox{R}_{\rm{Ti}}(\mu\to e)\gtrsim 10^{-17}-10^{-14}$ for the case with the RH neutrinos heavier than $10^9$\,GeV.
The LFVs originate from the slepton mass mixing, which is induced by the neutrino Yukawa interactions together with the large soft SUSY breaking mass for the up-type Higgs. 
We have confirmed that the degeneracy of the RH neutrino masses reduces the uncertainty in the relationship between the neutrino Yukawa couplings and the RH and the SM neutrino masses.  
Since the large soft SUSY breaking masses for the Higgs doublets seem to be inevitable to avoid the vacuum stability constraint in the stau-Higgs potential, this conclusion is somewhat model independent provided that the scale of SUSY breaking mediation is high enough.

 
\section*{Acknowledgments}
This work is supported by JSPS KAKENHI Grant Numbers JP16H06492 (N. Y.), JP16H06490, JP18H05542 and JP19K14701 (R. N.). The work of R. N. was supported by the University of Padua through the ``New Theoretical Tools to Look at the Invisible Universe'' project and by Istituto Nazionale di Fisica Nucleare (INFN) through the ``Theoretical Astroparticle Physics'' (TAsP) project.

\appendix
\section{Beta-functions for the slepton masses}\label{sec:app_a}
Here, we show one-loop beta-functions for the slepton masses. 
\begin{eqnarray}
 (16\pi^2)\frac{d m_{\tilde L}^2}{dt}&=& (m_{\tilde L}^2 + 2 m_{H_d}^2) Y_e^\dag Y_e 
 + 2 Y_e^\dag m_{\tilde E}^2 Y_e + Y_e^\dag Y_e m_{\tilde L}^2 + 2 A_e^2 Y_e^\dag Y_e \nonumber \\
 &+& (m_{\tilde L}^2+ 2 m_{H_u}^2 )Y_\nu^\dag Y_\nu 
 + 2 Y_\nu^\dag m_{\tilde N}^2 Y_\nu + Y_\nu^\dag Y_\nu m_{\tilde L}^2 
 +2 A_\nu^2 Y_\nu^\dag Y_\nu 
 \nonumber \\
 &-&6 g_2^2 |M_2|^2 -\frac{6}{5} g_1^2 |M_1|^2 -\frac{3}{5}g_1^2 S,\nonumber \\
 (16\pi^2)\frac{d m_{E}^2}{dt}&=& (2m_{\tilde E}^2 + 4 m_{H_d}^2) Y_e Y_e^\dag
 + 4 Y_e m_{\tilde L}^2 Y_e^\dag + 2 Y_e Y_e^\dag m_{\tilde E}^2 + 4 A_e^2 Y_e Y_e^\dag
 -\frac{24}{5} g_1^2 |M_1|^2 +\frac{6}{5}g_1^2 S ,\nonumber \\
 S &=& m_{H_u}^2-m_{H_d}^2 
 + {\rm Tr}\left[ m_{\tilde Q}^2 -m_{\tilde L}^2 -2 m_{\tilde U}^2 + m_{\tilde D}^2 + m_{\tilde E}^2\right]
\end{eqnarray}
where $t=\ln Q_{r} $ with $Q_r$ being the renormalization scale.
Here, $Y_e$, $Y_\nu$, $m_{\tilde L}^2$, $m_{\tilde E}^2$, $m_{\tilde Q}^2$, $m_{\tilde U}^2$ and $m_{\tilde D}^2$ are $3\times 3$ matrices. 

\bibliography{lfv_refs} 
\bibliographystyle{JHEP}

\end{document}